\documentclass[11pt,a4paper,useAMS,preprint]{article} 
\pdfoutput=1

\usepackage{jcappub}
\usepackage{multicol,url,times,graphicx,amsfonts,aas_macros,bbm,ulem} 
\usepackage{slashed}

\newcommand{\figureWidthOne}{0.6\textwidth}
\newcommand{\figureWidthTwo}{1.\textwidth}

\def\rebel{ReBEL~}

\def\Msol{~h^{-1} \rm{M}_{\odot}}

\newcommand{\refsec}[1]{section \ref{#1}}
\newcommand{\reftab}[1]{Table \ref{#1}}
\newcommand{\reffig}[1]{Figure \ref{#1}}

\newcommand{\eq}[1]{Eq. \eqref{#1}}

\def\beq{\begin{equation}}
\def\beqa{\begin{eqnarray}}
\def\beqaa{\begin{eqnarray*}}
\def\eeq{\end{equation}}
\def\eeqa{\end{eqnarray}}
\def\eeqaa{\end{eqnarray*}}

\def\frac#1#2{{\textstyle{#1\over #2}}}

\def\etal{{\sl et al.}}
\def\simlt{\stackrel{<}{{}_\sim}}
\def\simgt{\stackrel{>}{{}_\sim}}

\def\eps{\varepsilon}

\def\ga{\gamma}

\def\de{\delta}

\def\La{\Lambda}

\def\ln{{\textrm{ln}\,}}

\def\H0{H_0}

\def\hmpc{h^{-1}\,{\rm Mpc}}
\def\hkpc{h^{-1}\,{\rm kpc}}
\def\cdm{\ensuremath{{\rm CDM}}~}
\def\lcdm{\ensuremath{\La{\rm CDM}}~}
\def\difd{\textrm{d}}

\title{DM haloes in the fifth-force cosmology}

\author[a,b,c]{Wojciech A. Hellwing,\note{Corresponding author.}}
\author[d]{Marius Cautun,}
\author[e]{Alexander Knebe,}
\author[a,f]{Roman Juszkiewicz,}
\author[e]{and Steffen Knollmann}
\affiliation[a]{Institute of Astronomy, University of Zielona G\'ora, ul. Lubuska 2, 65-001 Zielona G\'ora, Poland}
\affiliation[b]{Interdisciplinary Center for Mathematical and Computational Modeling, University of Warsaw, ul. Pawi\'nskiego 5a, 02-106 Warsaw, Poland}
\affiliation[c]{Institute for Computational Cosmology, Department of Physics, Durham University, Science Laboratories, South Rd, Durham DH1 3LE, UK}
\affiliation[d]{Kapteyn Astronomical Institute, University of Groningen, P.O. Box 800, 9747 AV Groningen, The Netherlands}
\affiliation[e]{Grupo de Astrof\'isica, Departamento de F\'isica
  Te\'orica, M\'odulo C-8, Facultad de Ciencias, Universidad Aut\'onoma de Madrid, 28049 Cantoblanco, Madrid, Spain}
\affiliation[f]{Nicolaus Copernicus Astronomical Center, ul. Bartycka 18, 00-716 Warsaw, Poland}

\emailAdd{pchela@icm.edu.pl}
\emailAdd{cautun@astro.rug.nl}

\abstract{
We investigate how long-range scalar interactions affect the properties of dark matter haloes.
For doing so we employ the ReBEL model which implements an additional interaction between dark matter particles.
On the phenomenological level this is equivalent to a modification of gravity. We analyse the differences between
five ReBEL models and $\Lambda$CDM using a series of high resolution cosmological simulations. Emphasis is placed on
investigating how halo properties change in the presence of a fifth force.  We report that the density profile
of ReBEL haloes is well described by the NFW profile but with mean concentrations from $5\%$
to a few times higher than the standard $\Lambda$CDM value. We also find a slight increase of the halo spin for haloes more massive
than $5\times10^{11}\Msol$, reflecting a higher rotational support of those haloes due to scalar forces.
In addition, the dark matter haloes in our models are more spherical than their counterparts in $\Lambda$CDM.
The ReBEL haloes are also more virialised, with a large difference from $\Lambda$CDM for strong fifth forces and a much smaller change
for weak scalar interactions.

}


\arxivnumber{1111.7257}

\begin{document}

\maketitle

\section{Introduction}
The \lcdm model has proven capable of explaining a tremendous amount of observational data. In the era of precision cosmology, from both an observational
and modelling perspective, we are left with a more and more detailed picture of the evolution of the universe. This precisely measured history can be used
to impose constraints not only on the cosmological parameters (e.g. \cite{WMAP07,2004PhRvD..69j3501T,2006PhRvD..74l3507T,2005MNRAS.362..505C}) but also on
the nature of dark matter (DM) \cite{CDM1,CDM2,CDM3,CDM4,CDM5,CDM6}.

The strong relationship between DM characteristics and the emerging large scale structure (LSS) has been investigated in detail
(e.g. \cite{LSCDM1,LSCDM2,LSCDM3,LSCDM4,LSCDM5,LSCDM6,LSCDM7,LSCDM8,LSCDM9}). Most of the attention has focused on large scale linear structures and on
the properties of highly non-linear small-scale objects like haloes and galaxies. There still remains a considerable amount of unanswered questions with
respect to the physical nature of DM. Due to the great difficulties involved in the direct detection of DM particles in Earth-based experiments,
astrophysical observations still form the most important source of information on the nature and properties of DM particles.

The currently established \lcdm paradigm has some still unsolved problems whose solutions may reveal additional details about DM. On galaxy scales,
these answered questions involve the precise quantitative understanding of galaxies rotation, the central cusp of DM haloes and explaining the observed
rich population of thin disk dominated spiral galaxies that contrasts the \lcdm high merger and accretion rate at low redshifts (for an excellent discussion
refer to \cite{Rebel,rebel_nature}). On Megaparsec scales, the most interesting and well known problem involves the void phenomenon, strongly emphasised
by Peebles in \cite{PeeblesVoid}. It concerns the apparent discrepancy between the number of observed dwarf halos in voids and that expected from \lcdm simulations.

While there are clear discrepancies between observations and the predictions of \lcdm cosmological simulations, there is not a clear consensus if these
problems are due to our inability to simulate the universe in enough details or because of yet unknown physics present in the dark sector. This motivated
us to investigate the effects of additional DM physics on formation and evolution of structures in the cosmos. Studies of exotic new physics in the DM
sector are also important from a high-energy physics point of view. The results of modified DM models when compared with current cosmological observations
can be used to impose constraints and narrow down the possible DM candidates.

One of such possible \cdm paradigm modifications, formulated on the grounds of super-symmetry and string theory, has been proposed and developed by Gubser,
Peebles and Farrar \cite{GP1,GP2,PeeblesIDMDE,Farrar2007}. It consists of cold DM that, in addition to gravity, interacts by exchange of scalar particles.
This model, dubbed \rebel (\textit{daRk Breaking Equivalence principLe}), has shown the potential to solve some of the galaxy formation and evolution problems
discussed above. The difference between \rebel and the standard \cdm model consists in the additional interaction between DM particles, a so-called
``fifth force'', which acts exclusively on DM and has a limited range. Studies employing N-body simulations of the \rebel model
\cite{NGP,LRSI1,LRSI2,Rebel,Rebel2,reion1,LRSI3} and similar models of coupled DM and dark energy (DE) \cite{Coupled_DE1,Coupled_DE2,Coupled_DE5} have shown
that this class of models reproduces the observed large scale structure of the universe (e.g. the 1D power spectrum of the Ly-$\alpha$ forest galaxies or
the power spectrum from the SDSS galaxy survey\footnote{Sloan Digital Sky Survey - \url{http://www.sdss.org/}}). At the same time, it was shown that the
\rebel model introduces new effects, potentially positive, to the process of structure formation on galactic scales.

The recent years have seen several works that studied the effects of modified DM and DE models \cite{scalars1,scalars2,scalars3,scalars4,scalars5,
scalars6,scalars7,scalars8,scalars9,scalars10,scalars11,scalars12}. These have been motivated by recent observations posing even more challenges to the standard \lcdm
cosmology. Some of the new puzzles include an observed offset between baryonic and DM in clusters \cite{offset} and reports on very massive superclusters seen
at high redshifts \cite{UKIDSS_z09_supercluster, higz_cluster_obs1, higz_cluster_obs2}. These have important constraints and implications for both \lcdm and
modified DM and DE theories \cite{vectorDE,highz_clustersDE,highz_clustersLCDM}. The new observations have made the nature and properties of DM and DE a hotly
discussed and debated topic.

In this paper we study the impact of the \rebel model and its form of modified gravity on the internal properties of DM haloes. We do so by performing a suite of N-body simulations using \rebel DM physics that we compare with the results of the standard \lcdm cosmology.
This work comes into the international and multidisciplinary collective effort taken to further understand and study the nature of DM and its implication for cosmology.

This paper starts with the theoretical formulation and the physics of the \rebel model which are presented in \refsec{sec:therebel}. This is followed by
\refsec{sec:numerics} where we describe the numerical experiments that we performed to study the formation and evolution of DM haloes. The DM halo comparison
is presented in \refsec{sec:results_stat}.
Finally we end with \refsec{sec:conclusions} where we give our conclusions and final remarks.

\section{Scalar-interacting DM: the \rebel model}
\label{sec:therebel}

In this section we will briefly present, using \cite{GP1,GP2,NGP} as reference, the fundamental ideas and properties of the \rebel model discussed in this paper.

We consider a picture in which there exist an additional long-range force - different than gravity - that acts only on DM particles. Such a force arises due to
the interactions between DM particles and some underlying background scalar field $\phi$. Such an interaction can be expressed formally as:
\beqa
\label{eqn:action_dm1}
\mathcal{S} &=& -\int |\phi|\difd s\,,\quad \textrm{or}\\
\label{eqn:action_dm2}
\mathcal{S} &=& \int \sqrt{-g}\difd^4x(i\bar{\psi}\gamma\partial\psi - \phi\bar{\psi}\psi)\,.
\eeqa

The idea of long-range interactions appearing due to exchange of massless scalar particles has a very long history. Nordstr{\"o}m derived the classical form of
\eq{eqn:action_dm1} in 1913 \cite{Nordstrom1912}. This interaction is equivalent, in the limit of small de Broigle lengths, to the form given in \eq{eqn:action_dm2}
which was introduced by Yukawa \cite{Yukawa1935}.

Within the framework of quantum field theory there are arguments stating that it is very unlikely for any scalar field to escape from obtaining a large mass
($\gg H_0$), which would make any model with a scalar field particularly useless for cosmology. However, additional work on the grounds of the string theory
suggest that there is a possibility for a scalar field to maintain low mass.

In the beginning of the second half of the 20th century Pascual Jordan and Robert Dicke presented a few papers exploring the physics of the scalar-tensor gravity
implementing above action integral defined for particles in the Einstein frame\cite{Dicke1,Dicke2,Jordan}.

In 1990 Damour \etal \cite{DGB} noticed  that the tight empirical constraints that we have for long-range scalar interactions in the baryon sector allow for existence
of such interaction with significant strength in the dark sector. Modern considerations along this line of thought appear abundantly in the literature of this subject
\cite{GradwohlFrieman1,GradwohlFrieman2,skalar_teo1,skalar_teo2,skalar_teo3,skalar_teo4,skalar_teo5,skalar_teo6,skalar_teo7,skalar_teo8,skalar_teo9,skalar_teo10,skalar_teo11}.

We now focus on the model described in detail by Gubser and Peebels in \cite{GP2,GP1}. Let us consider at least two different species of DM particles which interact,
in addition to gravity, by a scalar field. This additional interaction is dynamically screened by the presence of light particles coupled by a Yukawa-like factor
to the scalar field. The generic Lagrangian for such a scenario is given by:
\beqa
		\label{eqn:lagrangian_dm}
		\mathcal{L} &=& \frac{1}{2}(\partial\phi)^2 +\bar{\Psi}_s i\slashed{\nabla}\Psi_s+\bar{\Psi}_+i\slashed{\nabla}\Psi_+\\
		&+& \bar{\Psi}_-i\slashed{\nabla}\Psi_- - y_s\phi\bar{\Psi}_s\Psi_s -(m_++y_+\phi)\bar{\Psi}_+\Psi_+ - (m_--y_-\phi)\bar{\Psi}_-\Psi_-\,,\nonumber
\eeqa
where $\slashed{\nabla}$ is the operator in Feynman slashed notation:
\beq
		\label{eqn:slashed_nabla}
		\slashed{A}\equiv\gamma^\mu A_\mu\,.
\eeq
Here $\ga^\mu$ are Dirac's gamma matrices and the equation is written using the Einstein summation convention. The constants $m_\pm$ and $y_\pm$ are both positive.
The fermions $\Psi_\pm$ are non-relativistic DM while the additional species of light particles $\Psi_s$ consist of the screening particles.

The action for the two particle species takes the form:
\beq
		\label{eqn:action_dm3}
		\mathcal{S}=\int\sqrt{-g}\difd^4x\phi_{,i}\phi^{,i}/2 - \sum_{\textrm{particles}}\int\left[m_+(\phi)\difd s_++m_-(\phi)\difd s_-\right]\,,
\eeq
where the symbol ${,i}$ denotes a partial derivative. The DM particles carry effective scalar charges $Q$:
\beq
		\label{eqn:scalar_charges}
		Q_+\equiv{\difd m_+\over \difd\phi}<0,\quad Q_-\equiv{\difd m_-\over \difd\phi}>0,\quad {\difd^2 m_\pm\over \difd^2\phi}\geq 0\,,
\eeq
which in general are conserved for non-relativistic phenomena\footnote{In general the scalar charges are not conserved for interactions at relativistic energies.
Additionally, the scalar charge of DM trapped in a black holes is lost.}. The field $\phi$ will undergo quasistatic relaxation towards equilibrium, yielding larger
values in the regions where the (+) particles dominate and smaller values in aggregations of the (-) particles. Thus, a particle with positive (+) charge will be
attracted towards regions with large $\phi$ values because it has a lower energy  $m_+(\phi)$. Similarly negatively charged particle (-) will  be attracted to
the regions with small $\phi$ values. This behaviour implies that particles with the same charge attract each other while particles of different kinds repel.

This basic scenario is similar to the so-called multi-coupled dark energy models (see e.g. \cite{multic-dm1,multic-dm2,multic-dm3,multic-dm4}). However, in this study we investigate the most basic and simple model of the \rebel cosmology, 
in which there is only one heavy species of scalar particles acting effectively as DM. The basic scheme that we assume admits for a relatively simple numerical implementation, that, at the same time,
 allows us to study a model
that can be understood as a generic fifth-force model. In this context, our simplest \rebel scenario is more generic than the phenomenologically rich multi-coupled dark energy models. Hence further we assume that there are two species of the DM particles which satisfy:
\beq
		\label{eqn:dm_mass}
		m_{DM} = m - y\phi,\quad m_s=y_s\phi, \quad\textrm{and}\,\, y\bar{n}<y_s\bar{n_s}\,
\eeq
where $m_{DM}$ is the mass of heavy particle species (effectively the mass of a DM particle), $m_s$ labels the mass of the screening particles, and
$\bar{n}$ \& $\bar{n_s}$ are the corresponding number densities of this particles. The scalar field $\phi$ has relaxed to quasistatic equilibrium, for which $m_s\sim 0$.
Thus, the screening particles are relativistic and generate the potential:
\beq
		\label{eqn:Vs}
		V_s = \sum_{\textrm{particles}}\int y_s\phi\difd s\simeq\int \difd^4ry_s\phi n_s\langle\sqrt{1-v^2}\rangle\,.
\eeq
This implies that:
\beq
		\label{eqn:Vs_dp}
		{\de V_s\over\de\phi}= y_sn_s\langle\sqrt{1-v^2}\rangle\simeq{y_s^2\bar{n_s}\over\eps_s}\phi, \quad\textrm{with:}\;\eps_s={y_s\phi\over\sqrt{1-v^2}}\,.
\eeq
Where the $\eps_s$ labels the averaged energy of a screening particle and $\langle\sqrt{1-v^2}\rangle$ is the mean velocity of the screening particles (when $c=1$).
The equation for the scalar field now takes the form:
\beq
		\label{eqn:equation_for_phi}
		\nabla^2\phi={\phi y_s^2\bar{n}_s\over\eps_s} - yn(\mathbf{r},t)\,.
\eeq
The above equation is consistent with an effective damping of the scalar field $\phi$ by the term:
\beq
		\label{eqn:e-screening}
		r_s=\sqrt{\eps_s\over y_s^2\bar{n}_s}\;\;\left[\textrm{Mpc}\right]
\eeq
which we call the screening length\footnote{This term is equivalent to the cut-off length in a Yukawa-like potential}. Making use of the $r_s$ definition, we rewrite
the scalar field equation as:
\beq
		\label{eqn:equation_for_phi2}
		\nabla^2\phi={\phi\over r_s^2} - yn(\mathbf{r},t)\,.
\eeq
Let us analyse this last equation in more detail. The last term describes the non-relativistic particles in a hydrodynamic approximation. The term $\phi/r_s^2$ appears
here \cite{PeeblesIDMDE}, because the source term of the $\phi$ field for a particle with velocity $v$ contains factor $ds/dt=\sqrt{1-v^2}$ and for quasistatic
configurations of the $\phi$ the energy of the screening particles $\eps_s$ (see \eq{eqn:Vs_dp}) is nearly independent of their location. Elimination of the $\sqrt{1-v^2}$
in favor of $\eps_s$ leads to the equation for the screening length (\ref{eqn:e-screening}). The $\eps_s$ energy does not depend on time since it is constant, thus due to
the expansion of the Universe it has to scale like $\eps_s\propto a(t)^{-1}$. Taking into account that $n_s\sim a(t)^{-3}$, we conclude that the screening length grows
like $r_s\sim a(t)$. Hence it is a constant in comoving coordinates.

At small distances ($r\ll r_s$) the scalar field generated by a single DM particle has the value $\phi=y/4\pi r$. Taking into account that the force exerted by this
field on another DM particle equals to the negative gradient of the mass $m-y\phi$, we obtain:
\beq
		\label{eqn:scalar_force1}
		\mathbf{F_s} = y\nabla\phi\,.
\eeq
Which clearly shows that two DM particles are attracted by the scalar interaction:
\beq
		\label{eqn:scalar_force2}
		F={y^2\over 4\pi r^2}\,\quad\textrm{for } r\ll r_s\,.
\eeq
Two DM particles also interact gravitationally. This leads us to define the $\beta$ parameter which is the ratio of the scalar to gravitational force strength between
two identical DM particles. This parameter is given by:
\beq
		\label{eqn:beta-scalar}
		\beta = {y^2\over 4\pi G m^2}\,\quad\textrm{for } r\ll r_s\,.
\eeq
Setting $\beta\sim \mathcal{O}(1)$ gives us scalar interactions which have a magnitude comparable to that of gravity. For distances much larger than the screening length
this scalar interactions fade away to zero. Hence the formula for the total force exerted on two DM particles can be splitted into two limiting cases:
\beqa
	\label{eqn:scalar-force-case1}
	F_{DM} = (1+\beta) F_{N}\quad\textrm{for } r\ll r_s\;,\\
	\label{eqn:scalar-force-case2}
	F_{DM} = F_{N}\quad\textrm{for } r\gg r_s\;.
\eeqa
To summarise, we are dealing with a model of DM which, in addition to gravity, interacts by means of a scalar field. The model introduces two free parameters $\beta$
and $r_s$, which are sufficient for a full phenomenological description.

\subsection{The phenomenological model and modified gravity}

Following the earlier work of \cite{NGP,LRSI1}, we study the \rebel model using its phenomenological parametrisation. In this approach we model the scalar interactions
appearing in the dark sector as effectively modified gravity. This is clear when we write the modified potential between two DM particles as:
\beq
		\label{eqn:rebel-pot}
		\Phi(\mathbf{r}) = -{Gm\over r}h(r)=\Phi_{N}\,h(r)\,
\eeq
with
\beq
		\label{eqn:rebel-h(x)}
		h(r) = 1+ \beta e^{-r/r_s}\,.
\eeq
Here $G$ is the Newton gravitational constant, $\mathbf{r}$ marks the separation vector between particles and $\Phi_N$ is the pure Newtonian potential. As we have shown
in the previous section, our model is described by two parameters: $\beta$ - the dimensionless factor measuring the ratio between scalar and gravitational forces
and $r_s$  - the screening length expressed in Mpc. Elementary considerations from string theory \cite{GP1,GP2,NGP} give crude estimations for the values of the parameters as:
\beq
		\label{eqn:rebel-parameters-1}
		\beta \sim \mathcal{O(}1)\,,\quad r_s\sim 1\textrm{Mpc}\,.
\eeq
The potential given in \eq{eqn:rebel-pot} gives rise to a modified force-law between two DM particles:
\beqa
		\label{eqn:rebel-force1}
		F_{DM} = -G{m^2\over r^2}\left[1+\beta(1+{r\over r_s})e^{-r\over r_s}\right]=F_N\cdot F_s(\beta,\ga)\,,\\
		\label{eqn:rebel-force2}
		\textrm{where: } \ga\equiv {r\over r_s}\,,\qquad F_s(\beta,\ga) = 1+\beta(1+\ga)e^{-\ga}\,.
\eeqa
Where the $F_s$ term measures the force deviations from the usual Newtonian gravity and $F_N$ marks the Newtonian force. We call $F_s$ the \rebel factor.
For $\beta = 0$ or $\ga\gg1$ we will get $F_s\rightarrow 1$, hence $F_{DM}\rightarrow F_{N}$ and we recover the standard Newton force law.

Equations (\ref{eqn:rebel-pot})-(\ref{eqn:rebel-force2}) yield a simple phenomenological description of the ReBEL addition to the CDM model.

\subsection{The growth of structures in the \rebel\ model}

\begin{figure}
		\centering
		\includegraphics[width=\figureWidthOne,angle=-90.0]{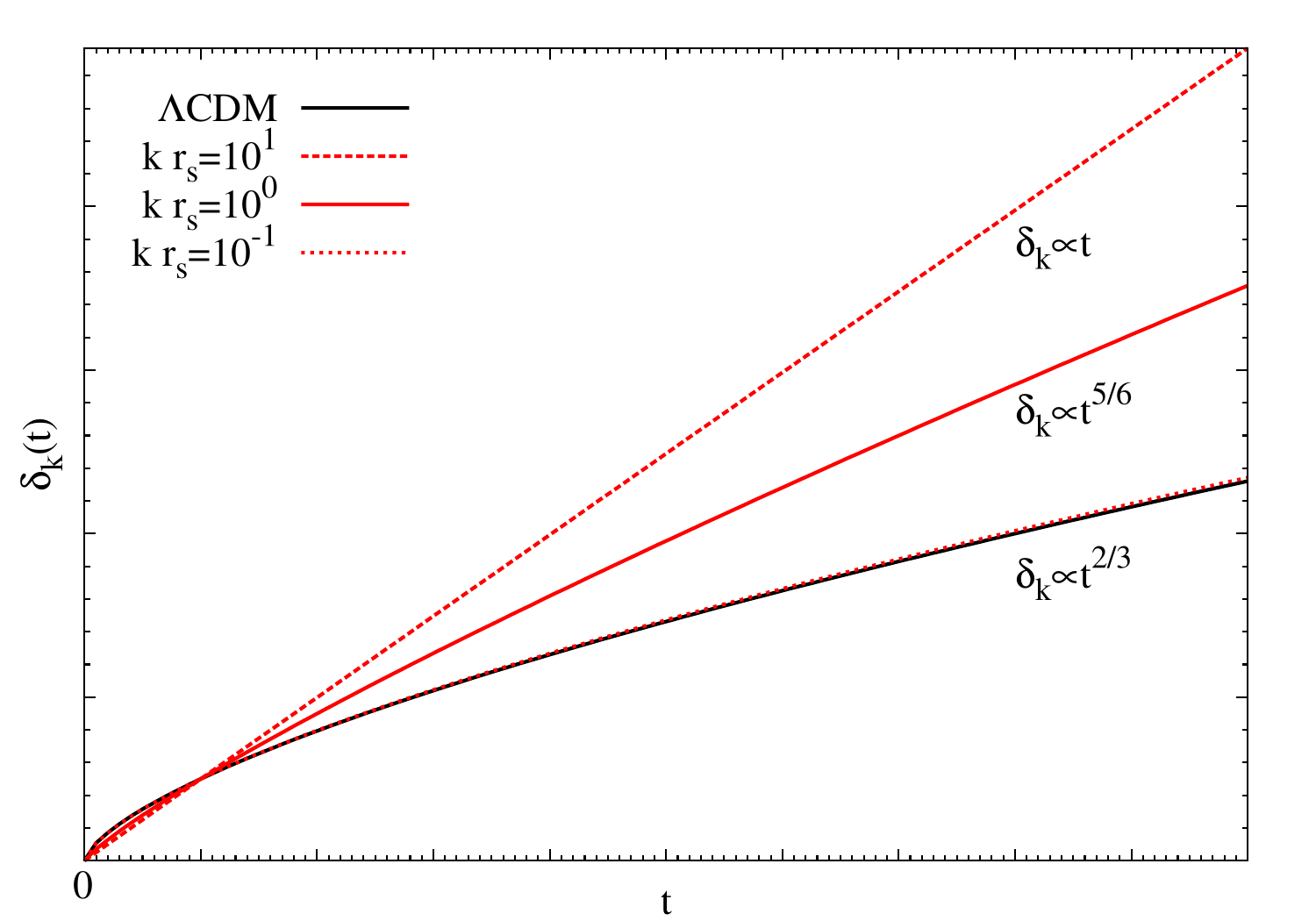}
		\caption[]{Growth of the Fourier modes of density perturbations in the linear regime during the matter dominated era as predicted by \eq{eqn:growth_mode}.
		Solid black line depicts the \lcdm model with $\delta_k\propto t^{2/3}$ for all values of $k$. Solid, dotted and dashed red lines mark the growth of
		density perturbations in the \rebel model for $\beta=1$ and at three different values of $r_s\cdot k$. Both axes have arbitrary units.}
		\label{fig:growth_compare}
\end{figure}

Now that we defined a phenomenological approximation to the \rebel DM physics (Eqs. \ref{eqn:rebel-pot}-\ref{eqn:rebel-force1}), we can analyse how the presence of
the scalar interactions affects structure formation. In doing so we limit ourselves to the matter dominated epoch since that period is especially important for galaxy
formation and evolution. We use linear perturbation theory to study the growth of density perturbations in the regime $\delta\simlt 1$. Because the \rebel model
changes the dynamics of DM only on scales small compared to the Hubble scale, \rebel does not affect the global evolution of the scale factor described by the Friedman
equations. Due to the modified force law given by \eq{eqn:rebel-force1}, the Poisson equation for the DM component of the \rebel model reads:
\beq
		\vec{\nabla}\times\Phi = 0\,,\quad \nabla^2_{\vec{r}}\Phi=4\pi G\rho\left(1+\beta e^{-|\vec{r}|/r_s}\right)
\eeq
To better understand the effects of the modified force, we linearise this equation together with the Euler and continuity equations in the limit of small perturbations
in $\delta$, $v_p$ and $\phi$ (the density, peculiar velocity and gravitational potential perturbations around a smooth background) \cite{1980Peebles}. The evolution
of the density perturbation for the matter-dominated epoch, which is well described by the Einstein-de Sitter universe with $\Omega=1$, is given by:
\beq
		\label{eqn:evol_rebel}
		\ddot{\delta_k}+{4\over 3t}\dot{\delta_k}={2\over 3t^2}\left[1+{\beta\over 1+(kr_s)^{-2}}\right]\delta_k
\eeq
which, for convenience, we write in Fourier space. The growing mode solution for this equation is \cite{NGP}
\beq
		\label{eqn:growth_mode}
		\delta_k\propto t^{\alpha}\,,\quad \alpha={1\over 6}\sqrt{25+{24\beta\over 1+(kr_s)^{-2}}}-{1\over 6}\,.
\eeq
For $k r_s\ll 1$ we get $\alpha = 2/3$, thus Fourier modes with $k\ll r_s^{-1}$ grow exactly in the same manner like in the standard \lcdm model. On the other hand
perturbations on scales smaller than characteristic screening length ($k\simgt r_s^{-1}$) grow faster than in the standard model. For example in the particular case with
$\beta=1$ and $k r_s\gg 1$ we have $\delta_k\propto t$. A comparison of the time evolution of Fourier density modes between \lcdm and \rebel is shown in \reffig{fig:growth_compare}.
The figure gives the growth of density fluctuations $\delta_k$ in a \rebel $\beta=1$ model for three modes satisfying $kr_s=10$, $1$ and $0.1$. The figure shows
that structure formation is enhanced and accelerated in \rebel compared to the \lcdm case on scales $\simlt r_s$. Moreover, different Fourier modes experience
different time evolution, from $\delta_k\propto t$ for $kr_s\ll1$ to $\delta_k\propto t^{2/3}$ for $kr_s\gg 1$. Faster growth of the small-scale density !
perturbations in ReBEL make these modes to cross into non-linear regime ($\delta\gg1$) quicker, which should result in earlier DM haloes formation and virialisation.
This was already shown to some extent in \cite{Rebel,Rebel2}.

\section{N-body simulations}
\label{sec:numerics}

\begin{table*}[ht]
		\centering
		\caption[The parameters describing our N-body simulations.]{The parameters describing our N-body simulations. The $\beta$ and $r_s\;[\hkpc]$ depict
		the values of the \rebel model free parameters used in a given simulation run, $L\;[\hmpc]$ is the box size, $z_{ic}$ is the redshift of the initial
		conditions, $\Omega_m$ and $\Omega_{\Lambda}$ are the dimensionless DM and DE density parameters (at $z=0$), $\sigma_8$ is the r.m.s. of the density
		fluctuations smoothed on $R=8\hmpc$, $h$ is the dimensionless Hubble parameter, $m_p$ is the mass of a single DM particle in units of $10^{8}h^{-1}M_{\odot}$,
		$\varepsilon\;[\hkpc]$ is the force resolution, $l\;[\hkpc]$ labels the mean inter-particle separation and $N_p$ gives the total number of DM particles.}
		\label{tab:sim-parameters}
		{\footnotesize
		\begin{tabular}{lccccccccccccc}
				\hline
				Simulation & $\beta$ & $r_s$ & $L$ & $z_{i}$ & $\Omega_m$ & $\Omega_{\Lambda}$ & $\sigma_8$ & h & $m_p$ & $\varepsilon$ & $l$ & $N_{p}$ \\
				\hline
				LCDM      & -   & -    & 32 & 1100 & 0.3 & 0.7 & 0.8 & 0.7 & 0.203 & 6 & 62.5 & $512^3$ \\
				B005RS500 & 0.05 & 500 & 32 & 1100 & 0.3 & 0.7 & 0.8 & 0.7 & 0.203 & 6 & 62.5 & $512^3$ \\
				B01RS500  & 0.1 & 500 & 32  & 1100 & 0.3 & 0.7 & 0.8 & 0.7 & 0.203 & 6 & 62.5 & $512^3$ \\
				B05RS500  & 0.5 & 500 & 32  & 1100 & 0.3 & 0.7 & 0.8 & 0.7 & 0.203 & 6 & 62.5 & $512^3$ \\
				B05RS1000 & 0.5 & 1000 & 32 & 1100 & 0.3 & 0.7 & 0.8 & 0.7 & 0.203 & 6 & 62.5 & $512^3$ \\
				B1RS1000  & 1.0 & 1000 & 32 & 1100 & 0.3 & 0.7 & 0.8 & 0.7 & 0.203 & 6 & 62.5 & $512^3$ \\
				\hline
		\end{tabular} }
\end{table*}

As mentioned earlier, string theory considerations provide a crude estimate for the values of the two free parameters: $\beta\sim 1$ and $r_s\sim 1\hmpc$ \cite{LRSI1,NGP}.
According to this, in this paper we explore four possible values of the $\beta = 0.05, 0.1, 0.5 \& 1.0$ and two values of the screening length parameter $r_s = 0.5 \& 1\hmpc$.
We label the corresponding simulations runs as \emph{LCDM, B005RS500, B01RS500, B05RS500, B05RS1000, B1RS1000}, where B stands for $\beta$ and RS marks the $r_s$ parameter.
The \emph{LCDM} run is the one with $\beta=0$ (i.e. no scalar forces) and corresponds to the standard cosmological model \lcdm.

All simulations were started with the same initial conditions at $z_{ic}=1100$. We need to adopt such a high starting redshift, since, in general, the scalar forces in
our scenario start acting on DM shortly after recombination. This set-up allows for a proper treatment of any non-linearities that may arise in early Universe due to
scalar forces.
Earlier \rebel studies have usually assumed $\beta\geq 0.5$, neglecting that scalar forces arise already in the early universe.
Therefore, one should understand the previous results of \cite{NGP, LRSI1, LRSI2, LRSI3,Rebel2} in a purely phenomenological way, where the value of the $\beta$
parameter describes just a toy-model approximation to the \rebel picture. Allowing the scalar forces to start acting at very high redshift (e.g. $z\approx1100$)
produces more pronounced effects at $z=0$. By studying these effects we found that models with with $\beta\geq 0.5$ and $r_s\geq500\hkpc$ values produce cosmic density fields characterised by a higher $\sigma_8(z=0)$ value than the fiducial \lcdm run. For example, the B1RS1000 model has a $\sigma_8$ value $10\%$ higher than the \lcdm cosmology. The higher variance of the density field, and the value of the $\sigma_8$ parameter in particular,
is the direct outcome of the fifth-force that enhances DM clustering. This effect was already shown and studied for the \rebel model \cite{LRSI1,LRSI3}
as well as for other modified gravity models (e.g. \cite{scalars12,fofr_2}.)

To follow the formation of structures within the \rebel framework we used an adapted version of the \verb#GADGET2# code \cite{Gadget2}. For the detailed descriptions of
the modifications made to the code we refer the reader to our previous paper on this subject \cite{LRSI1}. We conducted a series of high-resolutions DM only $N$-body
simulations containing $512^3$ particles within a periodic box of $32\hmpc$ comoving side length. We used the canonical \lcdm cosmology with $\Omega_m=0.3$,
$\Omega_{\Lambda}=0.7$, $\sigma_8=0.8$ and $h=0.7$.
Our simulation had a $m_p\simeq2.033\times 10^{7}h^{-1}M_{\odot}$ mass resolution and used a $\varepsilon=6\hkpc$ force softening parameter. In this study we limit our
analysis to the $z=0$ epoch, leaving the time evolution of the \rebel models as potential future work. Furthermore, the choice of a small simulation box size is motivated by our choice to have very high mass and force resolutions. This is done at the expense of a statistically relevant number of high-mass objects. However, in this work
we want to emphasize the fifth-force effects induced on galaxy and dwarf sized DM haloes. Hence, we tailored our simulations
for that purpose, leaving the high-mass and large-scale studies for future studies. The parameters of our simulations are summarised in \reftab{tab:sim-parameters}.

Before going further, we would like to make some remarks concerning our numerical setup. When choosing such a high initial redshift for our simulations, we must be careful as this can be connected with additional unwanted effects. The problems are two-fold: (i) First, at such a high redshift the density fluctuations have very small amplitudes and therefore
the N-body representation of the continuous density field is more prone to shot-noise effects related to the discrete nature of N-body particles. However, we
checked that the initial as well as intermediate times power spectra against shot-noise effects and found that the density perturbations are represented faithfully down to the Nyquist limit ($k_{Nyq}\sim 50 h/Mpc$) of our simulations. (ii) Secondly, as the GADGET2 code uses a finite-difference scheme for force interpolation,
small particle accelerations at early times are prone to a larger relative error in force estimation. It can affect our results if the relative error is much larger for the \rebel runs than in the LCDM case. To overcome this, we adapt a modified cell-opening criterion for
the \rebel runs, such that the relative error in the sum of the gravitational and fifth forces is consistent with the original (i.e. Newtonian) relative error
of the acceleration. This modifications is similar to the setup developed by Keselman \etal\cite{Rebel}, for a detailed discussion we refer the reader to the appendix
of \cite{Rebel}. Moreover, we employ a fine mesh size ($l_{cell}=31.25\hkpc$) for the PM part of the gravitational force calculation of GADGET2 that is much smaller than the
characteristic screening length $l_{cell}\ll r_s$. Thanks to this, the relative error in the estimation of the gravitational and fifth forces differs by an insignificant amount from the
usual Newtonian error (see Eqn. A2 in \cite{Rebel}). 

We used the MPI+OpenMP hybrid \verb#AMIGA# halo finder (\verb#AHF#), which is the successor of the \texttt{MHF} halo finder \cite{MHF}, to identify haloes and subhaloes
in our simulation\footnote{AHF is freely available from \url{http://popia.ft.uam.es/AMIGA}}. For a detailed description of \texttt{AHF} see the code paper \cite{AHF}.
We identify haloes using $R_{200}$ as halo edge. This radius is defined, as the distance at which the overdensity within a sphere centred at the halo centre satisfy:
\beq
	\overline{\rho}(r_{200}) \sim 200 \cdot \rho_{crit}\,,
\label{eqn:r200}
\eeq
where $\rho_{crit}$ is the critical density for a flat Universe. We take the $R_{200}$ and $M_{200}$ values as the virial radius and mass of a halo. It is important to
note that we needed to adjust the code to take into account the fifth force of the \rebel models. We have modified the unbinding procedure for removing the gravitationally
unbounded particles by using the modified potential laws of the \rebel model.
Moreover we also changed the halo circular velocity equation. We need to take into account that DM particles, due to the additional scalar force, have higher potential
energies in \rebel compared to \lcdm. The proper circular velocity equation at distance $R_{200}$ in the \rebel DM haloes has the following form:
\beqa
\label{eqn:vcir-rebel}
V_{c}^{ReBEL}&=&\left[{GM_{200}\over R_{200}}\cdot\left(1+\beta(1+R_{200}/r_s)e^{-R_{200}/r_s}\right)\right]^{1/2}=\nonumber\\
&=&V_{c}\cdot\left(F_s(\beta,R_{200}/r_s)\right)^{1/2}\,,
\eeqa
where $M_{200}$ is DM mass inside $R_{200}$, $V_c$ is the circular velocity for pure Newtonian dynamics and $F_s$ is the \rebel force factor given in \eq{eqn:rebel-force2}.
This expression can be found by noting that on a circular orbit the centripetal force acting on a DM particle has an additional component coming from the scalar interactions.

Throughout this paper we limit our analysis, if not explicitly stated otherwise, to DM haloes that contain at least $100$ DM particles. Thus we set our minimal halo
mass to $M_{min}=2.03\times 10^{9}h^{-1}M_{\odot}$.

\subsection{The effects of baryons}

The simulations used in this paper contain only a collisionless component. In other words, our approach
treats baryons as part of dark matter component, since the only baryonic effects that we consider are those incorporated in the initial transfer function (the BAO wiggles). 
Obviously, such an approach is a severe simplification of the real universe since it is well known that hydrodynamical and gas effects play a crucial role in galaxy formation \cite{bar1,bar2,bar3}.
At the same time, the presence of a baryonic component affects the DM distribution, with crucial effects
especially in the inner parts of haloes \cite{bar4,bar5,bar6}. This interplay between baryons and DM is further complicated in the \rebel scenario due to the breaking of the weak equivalence principle. As this
effect by itself is very interesting and leads to a more complicated picture (see e.g. \cite{equiv1,equiv2}), we choose to focus only on the net effects
induced by the fifth-force approximation on the DM distribution. We do so because the plethora of complicated
baryonic feedback processes (including effects such as radiative cooling, reionisation, supernova and AGN feedbacks)
would be very hard to disentangle from effects induced purely by the modified gravity. In fact, Puchwein \etal~\cite{mog_gadget}
 showed that in some class of fifth force models, baryonic feedback processes
can have opposite effects compared to the fifth force ones. Moreover, a proper modelling of galaxy-scale
baryonic processes depends on a suitable choice and tuning of sub-grid physics parameters.
Therefore, being fully aware of the limitation of our approach we decided to leave at a side the baryonic component for this current study.

\section{Results: Statistical properties of the halo populations}
\label{sec:results_stat}
In the subsequent analysis we focus on characterises how the presence of the long-range scalar interactions affect the whole population of DM haloes. In doing so we compare
the differences between the fiducial \lcdm cosmology and the \rebel models on the basis of changes in the distribution and statistical means of halo properties like density
profile, spin and halo shape.

\subsection{Density profiles}
\begin{figure}
    \centering
	\includegraphics[width=\figureWidthOne,angle=-90]{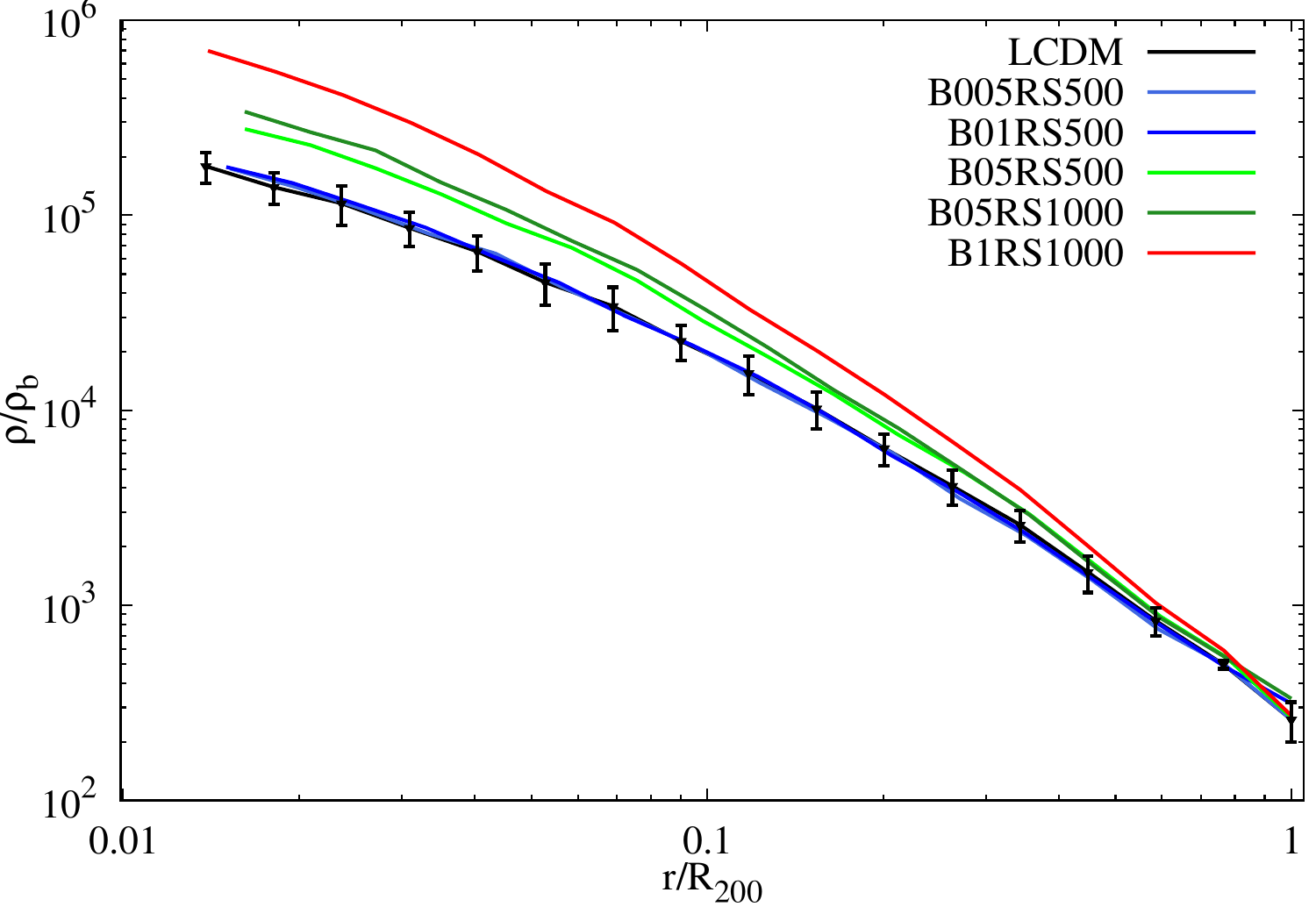}
    \caption[]{The spherically averaged density profiles for the 10 Most Massive Haloes of the \lcdm and \rebel runs.}
    \label{fig:profiles_MMH}
\end{figure}

Our phenomenological \rebel models effectively change the strength of gravity on scales $r\leq r_s$, thus we expect to see changes in the internal structure of DM haloes.
One way to quantify these changes is via the halo density profile.
To get good statistics and eliminate the halo to halo variation, we compare the density profiles obtained from averaging over the 10 Most Massive Haloes (MMH10) found in each
of our simulations. These haloes have masses in range $10^{13}M_{\odot}h^{-1} < M_{200} \simlt 10^{14}M_{\odot}h^{-1}$. We focus only on the most massive haloes since they
have virial radii larger than $\approx 400\hkpc$.
In the B005RS500, B01RS500 and B05RS500 models the size of these massive haloes is comparable to the screening length of the fifth force, whereas for B05RS1000 and B1RS1000
the screening length is larger than the virial radii. Hence we suspect that any significant deviations in the density profiles due to the \rebel model should be imprinted
in these massive haloes. The averaged density profiles for all the runs are shown in \reffig{fig:profiles_MMH}. We find that in the inner parts of the MMH10 ($r<0.1R_{200}$)
the \rebel haloes have higher density than the fiducial objects from the LCDM run. Thus we can expect that the counter-partners of the LCDM MMH10 have higher concentrations.
To asses this and better understand the density differences between models we use the well known  Navarro, Frenk \& White (NFW)
profile \cite{NFW}. This was shown to be a universally good fit for the majority of \lcdm haloes.
The NFW profile is given as:
\beq
	\label{eqn:NFW}
	\rho_{NFW}(R)={\rho_0\over (R/R_s)(1+R/R_s)^2}\,.
\eeq
Here $\rho_0$ and $R_s$ (not to be mistaken with $r_s$ - the scalar force screening length parameter) are the parameters of the density profile fit. The first parameter is
the characteristic density and the second one is the scaling radius. The scaling radius is usually used to define the concentration parameter $c_{200}$ of a given $NFW$
density profile:
\beq
	\label{eqn:con_param}
	R_{200} = c_{200}\cdot R_s\,,
\eeq
where $R_{200}$ is the virial radius of the halo.

\begin{figure}
	\centering
	\includegraphics[width=\figureWidthOne,angle=-90]{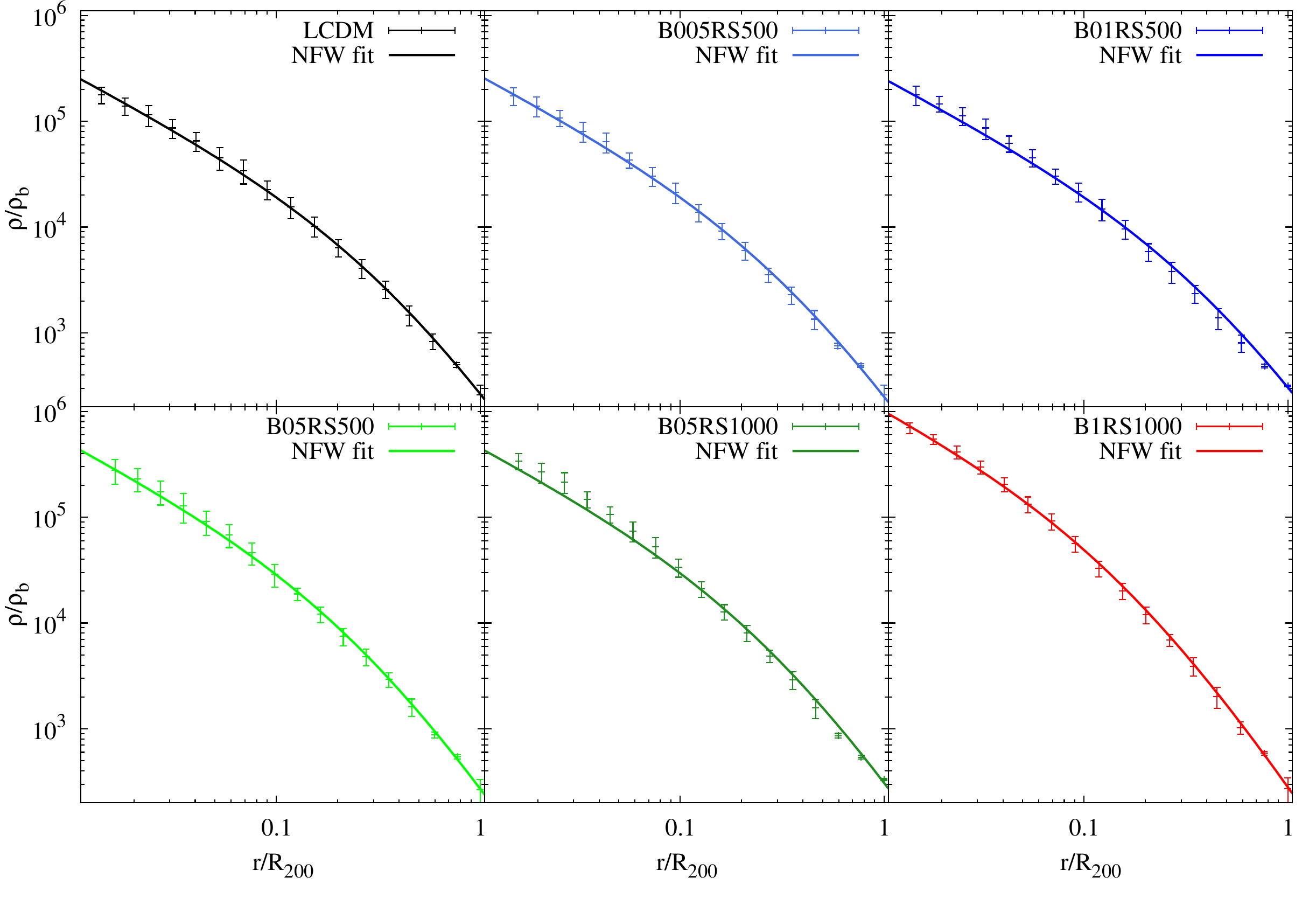}
    \caption[]{The MMH10 density profiles shown separately, together with the best NFW fits.}
    \label{fig:profiles_fits}
\end{figure}

The NFW fits to each of the MMH10 density profiles are shown in \reffig{fig:profiles_fits}. Each panel in the figure shows the profile and the fit for the 10 most massive
haloes in that simulation run.
We find that the NFW profile is a reasonably good fit for all our massive haloes, with small deviations only for some models below $0.1\;R_{200}$. These deviations are not
surprising since it is well established that the NFW profile is poorly describing very massive haloes that, in general, are not that well relaxed \cite{Wechsler2002}.
Thus we conclude the universality of \cdm halo density profiles even in the presence of a fifth force with different screening lengths. The parameters of the NFW fits
are summarised in \reftab{tab:NFWfit}.
\begin{table*}[ht]
    \centering
    \caption{The NFW best fit parameters for MMH10 density profiles. }
    \label{tab:NFWfit}
    \begin{tabular}{l|ccc}
        \hline
        Model & $\rho_0$ & $R_s/R_{200}$ & $c_{200}$\\
        \hline
        LCDM & 6624 & 0.4343 & 2.303\\
        B005RS500 & 7133 & 0.4111 & 2.433\\
        B01RS500 & 5499 & 0.4988 & 2.005\\
        B05RS500 & 16772 & 0.3008 & 3.324\\
        B05RS1000 & 15153 & 0.3299 & 3.031\\
        B1RS1000 & 63885 & 0.1831 & 5.463\\
        \hline
    \end{tabular}
\end{table*}
We find that the most massive haloes in the \rebel runs are characterised by higher characteristic densities and smaller scaling radii which in turn mean higher virial
concentrations. We confirm such findings for all \rebel MMH10s except the B01RS500 case. For this model the best NFW fit is characterised actually by lower concentration
than the LCDM case.

\begin{figure}
    \centering
	\includegraphics[angle=-90,width=0.8\textwidth]{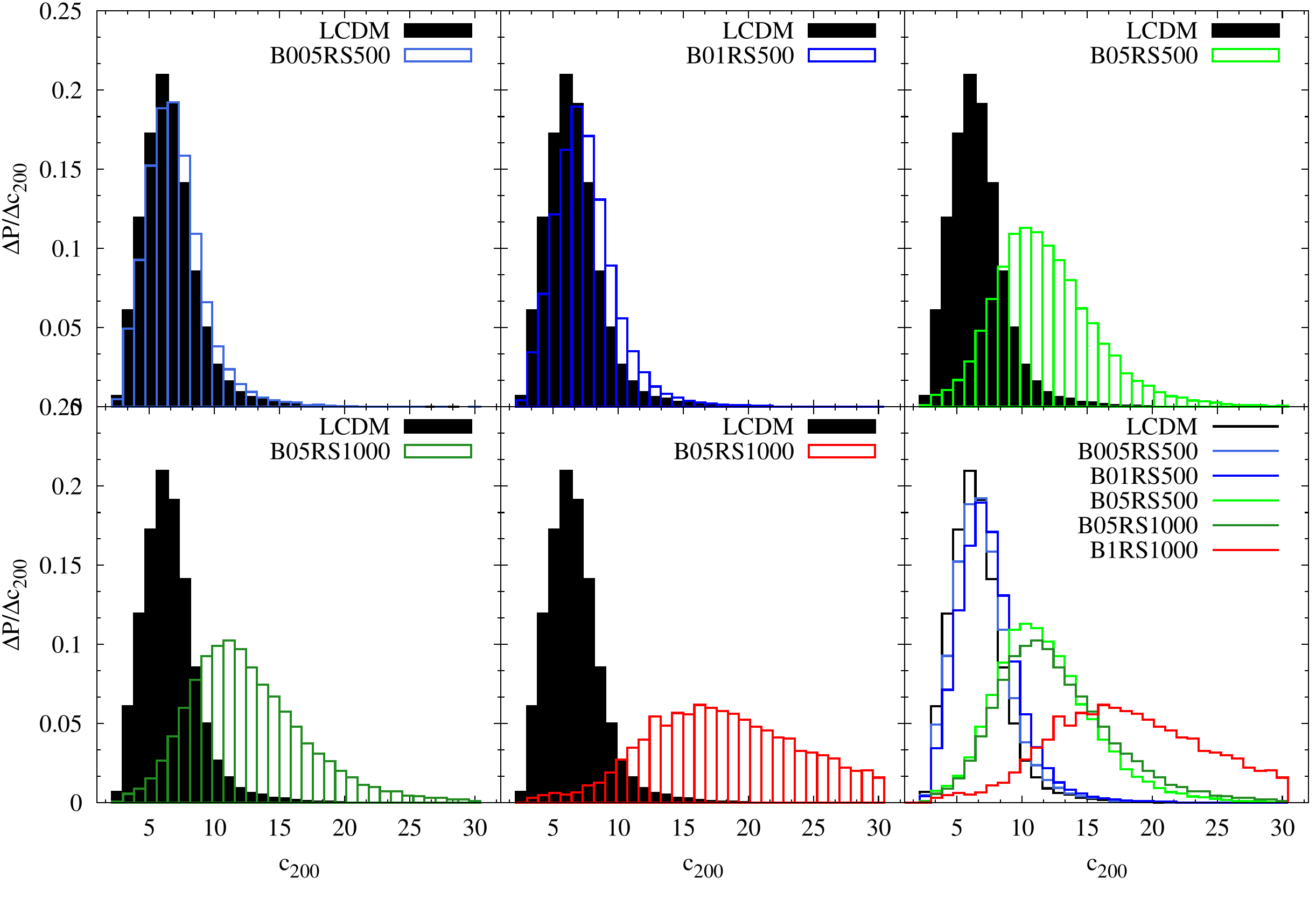}
    \caption[]{The distributions of the virial concentration parameter for all haloes in our \rebel models. Each panel shows also the fiducial \lcdm case for comparison. }
    \label{fig:cvir_dist}
\end{figure}
We find that the NFW fit is universally good in describing the density profile for the majority of haloes in the \rebel runs, not only for the most massive ones. Fitting
the density profile with a NFW function results in a concentration parameter $c_{200}$ for each halo. This $c_{200}$ parameter is a measure of the imprint of the fifth
force onto the haloes inner structure. The distribution of the concentration parameter for each run, for all the haloes, is shown in \reffig{fig:cvir_dist}. In each panel,
on top of the respective \rebel run, we plot with black boxes the distribution obtained from the LCDM run which is the fiducial case to compare with. The bottom-right
panel presents distributions for all runs to allow for a direct comparison between \rebel models. We find that the \rebel haloes are characterised by broader distributions
with higher values of the mean $c_{200}$ parameter. These findings are summarized in \reftab{tab:cvir_dist}.
\begin{table*}[ht]
    \centering
    \caption{The mean and standard deviation of the $c_{200}$ distributions when fitting NFW profiles to the whole halo population.}
    \label{tab:cvir_dist}
    \begin{tabular}{l|cc}
        \hline
        Model & $\overline{c_{200}}$ & $\sigma_{c_{200}}$ \\
        \hline
        LCDM & 5.9 & 2.7\\
        B005RS500 & 6.3 & 2.7\\
        B01RS500 & 6.81 & 2.8\\
        B05RS500 & 11.32 & 4.2\\
        B05RS1000 & 12.15 & 4.9\\
        B1RS1000 & 19.24 & 8.0\\
        \hline
    \end{tabular}
\end{table*}
\begin{figure}
    \centering
	\includegraphics[angle=-90,width=0.8\textwidth]{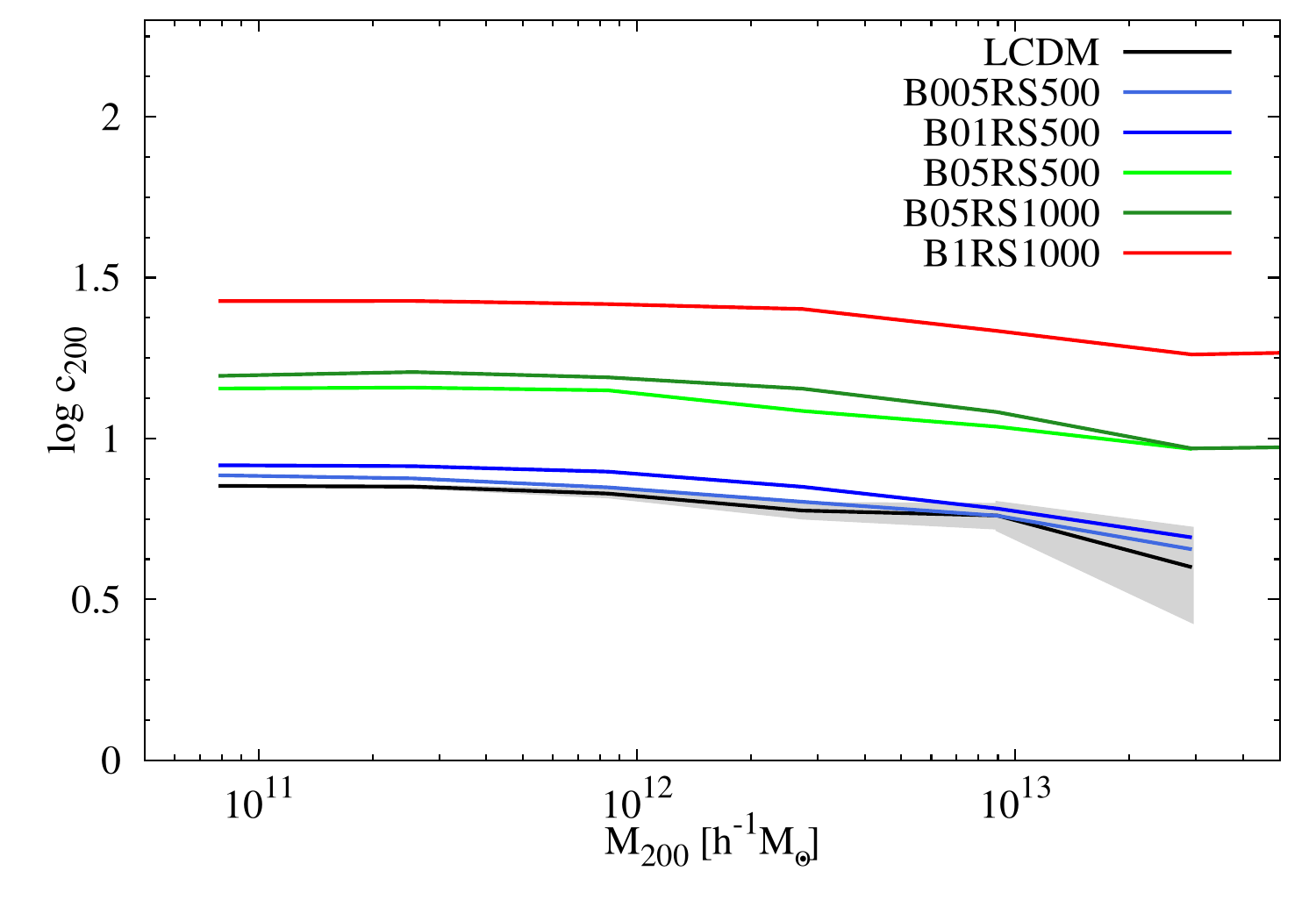}
    \caption[]{The mass-concentration relation. The median $c_{200}$ binned in halo mass is plotted for all models. The shaded region marks the bootstrap error
    of the median for the \lcdm case.}
    \label{fig:cvir_mass}
\end{figure}

Thus, on average, the \rebel haloes have higher concentrations. This effect is small only when we consider more realistic models with small $\beta$ value. For B005RS500
the mean concentration is only $6.7\%$ higher then in the LCDM case. For the B01RS500 model, the deviation reaches already $15.4\%$. When we look at the remaining models
with $\beta\geq0.5$ the effects become dramatic with the increase of the mean concentration $\overline{c_{200}}$ from $92\%$ for B05RS500 all the way to $300\%$ for B1RS1000.
The mean concentration for strong fifth force models shows a significant difference from the standard model, with values that are $2$ to $5\sigma$ higher than the \lcdm mean. However the distributions plotted in \reffig{fig:cvir_dist} are dominated by low mass objects, which biases our result to small haloes.
We investigate how these findings change as a function of halo mass in \reffig{fig:cvir_mass}.
The plot gives the median $c_{200}$ binned in halo mass for our models. The figure also shows the $1\sigma$ bootstrap error in the determination of the median for \lcdm which quantifies how significant is the deviation  given our limited halo sample.
It is clearly seen that the median concentration is systematically shifted towards higher values for all \rebel models and for all halo masses.

This can potentially mean bad news for the \rebel cosmology. Such a noticeable increase of the mean concentration poses additional challenges to
the problem of the bright dwarf satellites of the Milky Way galaxy \cite{lcdm_mw1,lcdm_mw2,lcdm_mw3}.
But we must admit that the full self-consistent quantification of this problem requires simulations in a
bigger box (to obtain a larger sample of Milky Way-like objects) that additionally must also contain a baryon component with full hydrodynamical solution.
Ergo we will not discuss this problem any more in this work.

Higher concentrations of the \rebel haloes are the outcome of boosted hierarchical structure formation in this class of models. This induces earlier formation times for
the majority of haloes as reported by Hellwing et al.\cite{Rebel2}, which potentially can be desired to solve some of the challenges of  \lcdm
(see the introduction for a discussion of such problems). But this is a double-edged sword that comes at the price of raised halo concentrations which poses additional difficulties.

\subsection{Spin \& halo rotation}
\label{sec:spin}
According to the tidal torque theory the angular momentum of a galaxy/DM halo is growing thanks to the gravitational tidal interactions between the DM proto-halo and
the matter distribution surrounding it \cite{amom1,amom2,amom3,amom4,amom5,amom6}. Thus we can express the angular momentum gained by the halo as
\beq
	\label{eqn:ang_momentum}
	\mathbf{J}_i\propto\varepsilon_{ijk}\mathbf{I}_{jl}\mathbf{T}_{lk}\,,
\eeq
where $\mathbf{I}_{jl}$ is the inertia tensor of the proto-halo and $\mathbf{T}_{lk}$ is tidal torque tensor generated by the matter distribution in the vicinity of the halo.
These tidal torque forces are also shaping the characteristic web-like pattern of the matter distribution on large-scale, hence providing a connection between the angular
momenta of DM haloes and the matter distribution on Megaparsec scales. However, the late-time dynamical evolution imprints non-linear and gasodynamical effects that
strongly erase primordial connection between the matter distributions and angular momenta of DM haloes. In hierarchical structure formation scenarios the highly non-linear
processes like tidal stripping, close encounters and major mergers dominate the growth of the angular momentum at intermediate and small redshifts (see details in e.g. \cite{Angular1,Angular2,Angular3,Angular4,Angular5}).
Nevertheless, knowing that in \rebel cosmologies the attractive forces shaping the evolution of DM are enhanced on Megaparsec scales,
we expect to see imprints of the scalar fields in the angular momentum of haloes.

\subsubsection*{The spin parameter}

The angular momentum of a DM halo is commonly parametrised by the dimensionless spin parameter \cite{amom2}
\beq
	\label{eqn:spin_peeb}
	\lambda={|J||E|^{1/2}\over GM^{5/2}}\,,
\eeq
where $E$ is the total mechanical energy of the halo, $M$ is its mass, $G$ is Newton's gravitational constant and $J$ is the total angular momentum.
There is another common definition of the dimensionless spin parameter proposed by Bullock \etal\ \cite{Bullock2001}. This second definition is more convenient to
implement computationally and we decided to use it in our study. The Bullock spin parameter is defined as
\beq
	\label{eqn:spin_bul}
	\lambda={|j|\over\sqrt{2}R_{200}V_{c}}\,,
\eeq
where
\beq
	\label{eqn:vcirc1}
	V_{c}=\left({GM_{200}\over R_{200}}\right)^{1/2}
\eeq
is the circular orbit velocity at $R_{200}$ and $j$ is the specific angular momentum of the halo
\beq
	\label{eqn:spin-spec}
	j={1\over N_H}\sum_{i=0}^{N_H}r_i\times v_i\,.
\eeq
Here the sum covers all particles belonging to a given DM halo $0\leq i<N_H$. Equation (\ref{eqn:vcirc1}) is correct for Newtonian gravity, but for \rebel haloes we
have to use the modified circular orbit velocity given by  \eq{eqn:vcir-rebel}. The spin parameter as defined above measures the rotational support of a halo.
High values of $\lambda$ corresponds to haloes that are spinning fast, while a low $\lambda$ indicates a slowly rotating halo.
\begin{figure}
    \centering
	\includegraphics[width=\figureWidthOne,angle=-90]{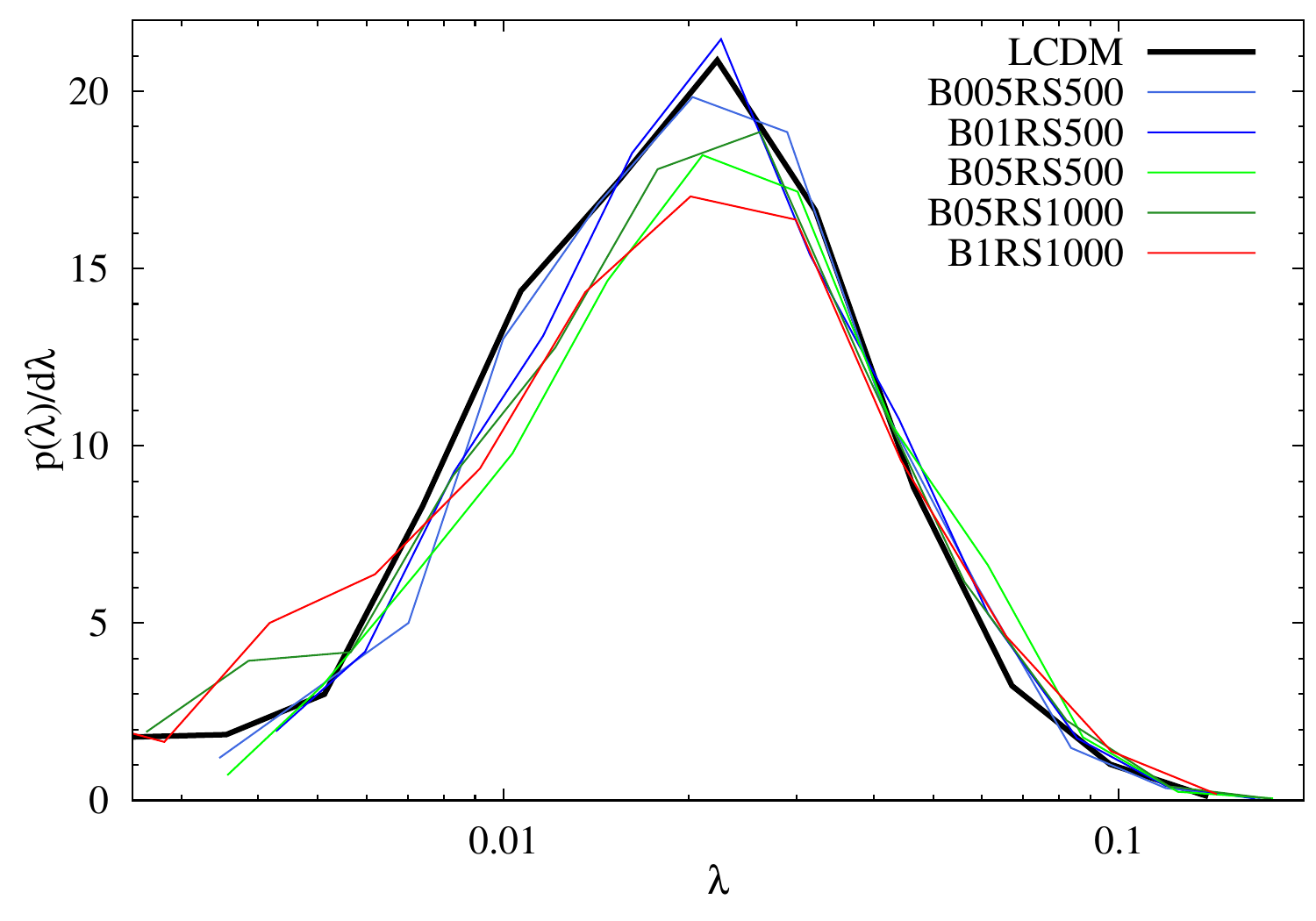}
    \caption[]{Distributions of the spin parameter $\lambda$ for our models.}
    \label{fig:spin_all}
\end{figure}

Our halo catalogues are dominated by low-mass objects with $M_{200}\leq 2\times 10^{10}M_{\odot}/h$. We have checked that for these dwarf-like objects the differences
in spin parameter distributions between \lcdm and \rebel are very small. The strongest difference was noted for the B05RS500 model, but that accounted to less than
$\approx 5.2\%$ of the mean \lcdm spin parameter.
Therefore in the following analysis we focus on bigger galaxy-like haloes with $M_{200}>10^{11}M_{\odot}/h$. For this population
of objects the deviations from the fiducial model are more prominent.

\reffig{fig:spin_all} gives the spin parameter distributions for our six simulations. We observe that there are clear differences between the \lcdm and \rebel spin
distributions. The fifth force models have a broadening and flattening of the distribution, with more prominent high and low spin tails. There also seem to be a
shift of the distribution centre towards higher $\lambda$ values when compared to the \lcdm case.
Therefore we conclude that haloes in the \rebel cosmologies have,
in a statistical sense, higher spin values, thus are spinning faster than \lcdm haloes. We note that the largest deviations from \lcdm appear for the strongest
\rebel models studied in this work, i.e. for B05RS1000 and B1RS1000. The shift of the spin distribution centres -- however noticeable -- does not appear to be large.
To check the statistical significance of this effect we use that the spin parameter distribution in N-body simulations is well approximated by a log-normal function given by
\beq
	\label{eqn:lognorm-p}
	p(\lambda)\textrm{d}\lambda = {1\over\lambda\sigma\sqrt{2\pi}}\exp\left[-{(\ln \lambda-\mu)^2\over 2\sigma^2}\right]\textrm{d}\lambda\,,
\eeq
with the centre of the distribution in the range $0.04<\lambda <0.05$ \cite{spin1,spin2,spin3,spin4,spin5}.
\begin{figure}
    \centering
	\includegraphics[width=\figureWidthOne,angle=-90]{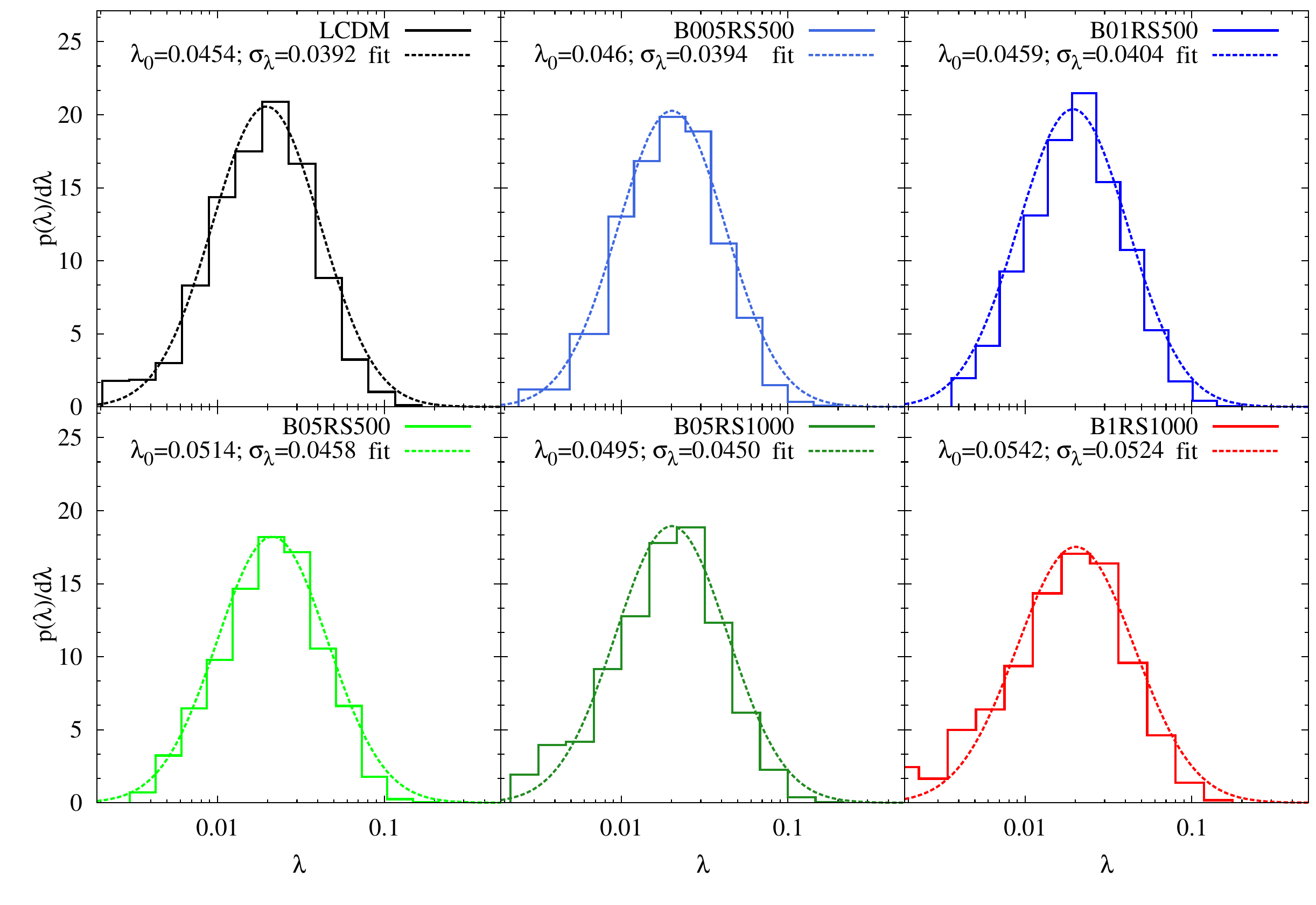}
    \caption[]{The distributions of the spin parameter accompanied by the fitted log-normal distributions. In each panel we also give the numerically computed mean
    value and the standard deviation of the shown distribution.}
    \label{fig:spin_dist}
\end{figure}
The log-normal distribution has the mean value (i.e. first moment) given by
\beq
	\label{eqn:lognor-mean}
	\lambda_0=e^{\mu+{1\over2}\sigma^2}\,,
\eeq
and the standard deviation as
\beq
	\label{eqn:lognor-sigma}
	\sigma_\lambda=e^{\mu+{1\over2}\sigma^2}\sqrt{e^{\sigma^2}-1}\,.
\eeq
\begin{table*}[ht]
    \centering
    \caption{The mean and standard deviation for the best log-normal fit to the spin distributions. }
    \label{tab:spin_fit_params}
    \begin{tabular}{lcc}
        \hline
        Model & $\lambda_0$ & $\sigma_\lambda$ \\
        \hline
        LCDM      & 0.0454 & 0.0392\\
        B005RS500 & 0.0460 & 0.0394\\
        B01RS500  & 0.0459 & 0.0404\\
        B05RS500  & 0.0514 & 0.0458\\
        B05RS1000 & 0.0495 & 0.0450\\
        B1RS1000  & 0.0542 & 0.0524\\
        \hline
    \end{tabular}
\end{table*}

In \reffig{fig:spin_dist} we plot the distributions of the $\lambda$ parameter and their corresponding best-fit log-normal functions as given by \eq{eqn:lognorm-p}.
Each panel shows the spin distribution and log-normal fit for that given model. The best fit log-normal functions describe reasonably well the underlying distributions
for all models -- even for the strongest \rebel run. The parameters of these fits are presented in \reftab{tab:spin_fit_params}. The mean $\lambda_0$ and standard
deviation $\sigma_\lambda$ of the log-normal fit are gradually increasing as we move from \lcdm to the strongest \rebel model B1RS1000, in complete agreement with
the qualitative observations we made for \reffig{fig:spin_dist}. The maximum change in $\lambda_0$ appears between the \lcdm model and B1RS1000 and amounts to over $19\%$.
Although this increase in $\lambda_0$ is not large, it is accompanied by the more dramatic broadening of the distributions. The increase in $\sigma_\lambda$ due to
the presence of the scalar forces reaches nearly $33\%$ when comparing \lcdm and B1RS1000. Therefore, on average, there are more fast rotating haloes in \rebel than
in the standard \lcdm cosmology. This can be a potentially positive effect, as fast rotating DM haloes promote easier thin-disk spiral galaxy assembly \cite{disk_halo1,disk_halo2}.

\begin{figure}
    \centering
	\includegraphics[width=\figureWidthOne,angle=-90]{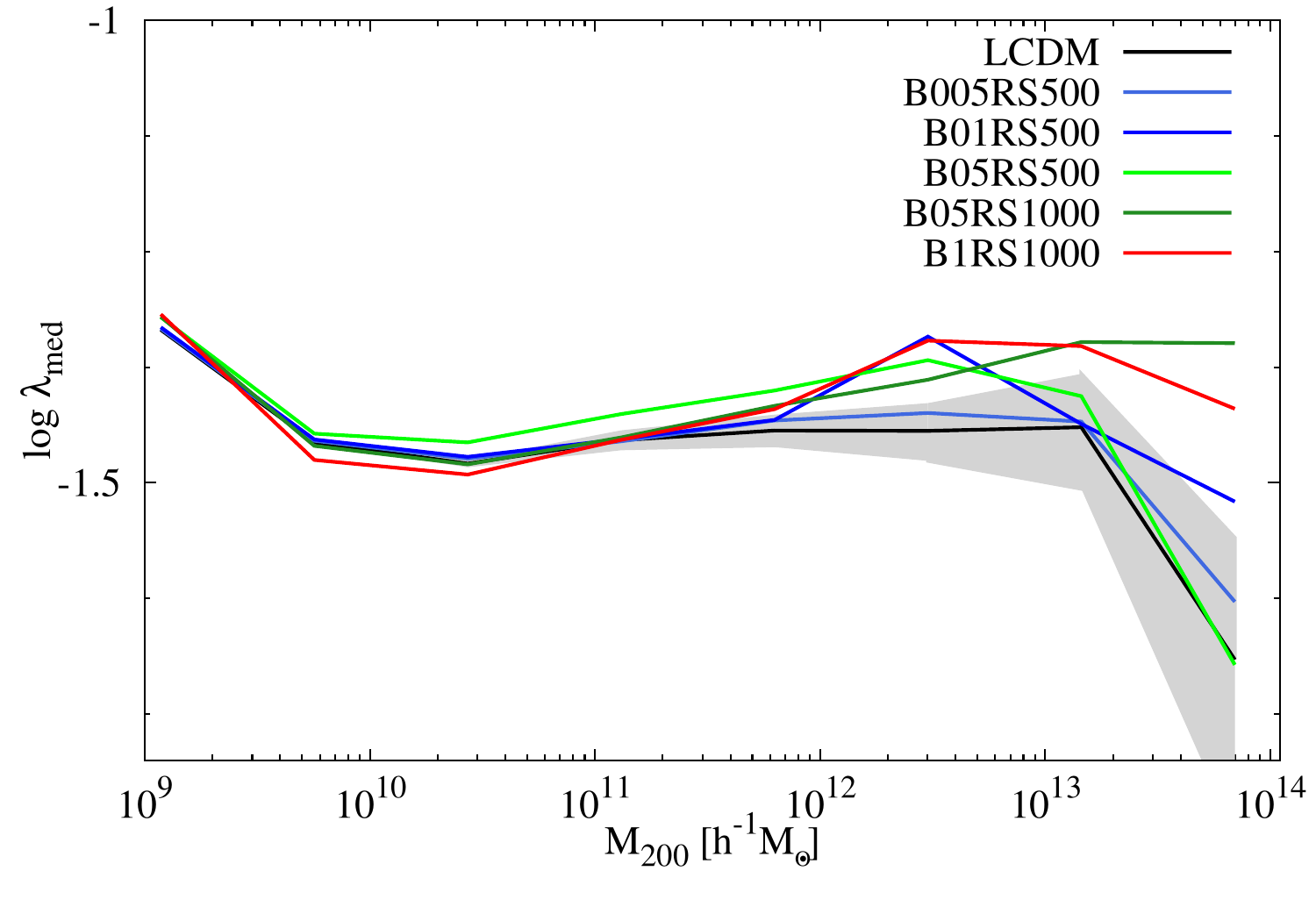}
    \caption[]{The $\lambda-M_{200}$ correlation for all our models. The shaded region reflects the error in the determination of the median $\lambda$.
    For clarity the errors are plotted only for the \lcdm case.}
    \label{fig:spin_mass}
\end{figure}

Earlier studies of the halo spin in \lcdm N-body simulations found that there is a weak dependence of $\lambda$ on halo mass \cite{spin_mass1,spin_mass2,spin_mass3}.
The observed dependence shows that more massive haloes tend to have slightly lower spin parameters.
The spin-mass dependence for \lcdm and \rebel is shown in \reffig{fig:spin_mass} where we plot the median spin parameter as a function of the halo mass. The error bars, plotted only for
the \lcdm case, represent the first and the last quadrille over the median. For \lcdm we observe that there is a marginal dependence on mass of the form
$\lambda\propto M^{\alpha}$ with the mean slope $\alpha=-0.005$. This is in agreement with other halo spin studies \cite{spin_mass1,spin_mass3}. When it comes to
the \rebel results, we find deviations from the fiducial \lcdm behaviour. The average spin of low-mass haloes is affected only very weakly by the modified DM
gravity. On the other hand, for massive objects we observe a strong increase of the mean spin compared to the \lcdm case. The presence of a fifth force breaks
the weak spin-mass dependence because it affects more strongly the massive haloes which tend to spin faster than the low mass one. Our simulations are not suited
to quantify the spin-mass relation for \rebel since we have only few objects at masses $M\geq 10^{12} M_{\odot}/h$ that give most of the variation with mass.
A detailed analysis of the \rebel effects on the spin-mass relation and the spin of cluster-like haloes needs a much better sample of massive objects that can be
obtained only in simulations with a larger volume. Hence we leave this subject for a future study.

It is interesting to note that for all \rebel models -- except the runs with $r_s=1000\hkpc$ -- the $\lambda_{med}$ values approach the \lcdm limit for
$M_{200}>10^{13}h^{-1}M_{\odot}$. Using that the mass of a halo corresponds to some length scale that is related to the virial radius $R_{200}$, the above result
implies that the halo spin for objects with large $R_{200}/r_s$ ratios is less sensitive to the presence of a fifth force. These observations agree with the tidal
torque theory of halo's angular momentum and spin. The scale of the tidal forces shaping the halo spin is related to a halo's virial radius. Thus for big haloes
the enhancement of forces on scales smaller than it's radius do not contribute significantly to the growth of its angular momentum.

\subsection{Shapes and geometry}
\label{sec:shapes}
The shape and the geometry of a halo are determined jointly by linear and non-linear processes. Theoretical predictions are usually constrained to the linear processes
influencing the shape of a DM halo. The high complexity characterising the non-linear phenomena acting within a halo makes predicting the direct outcome of such
processes very difficult. Generally one assumes that three principal factors are responsible for the shape of an halo:
\begin{itemize}
	 \item the shape and orientation of the primordial density peak from which the halo originated \cite{geo1,geo2,geo3,geo4},
	 \item the external tidal shear field that shapes the halo \cite{geo5,geo6},
	 \item and the non-linear interactions disturbing the original halo shape, e.g. the violent relaxation and halo merging \cite{geo7}.
\end{itemize}
The complex interplay of the above factors determines the final halo shape and geometry. In the following we check to what extent the presence of additional DM
scalar-interactions affects the shape of the DM halo.
\begin{figure}
    \centering
	\includegraphics[width=\figureWidthOne,angle=-90]{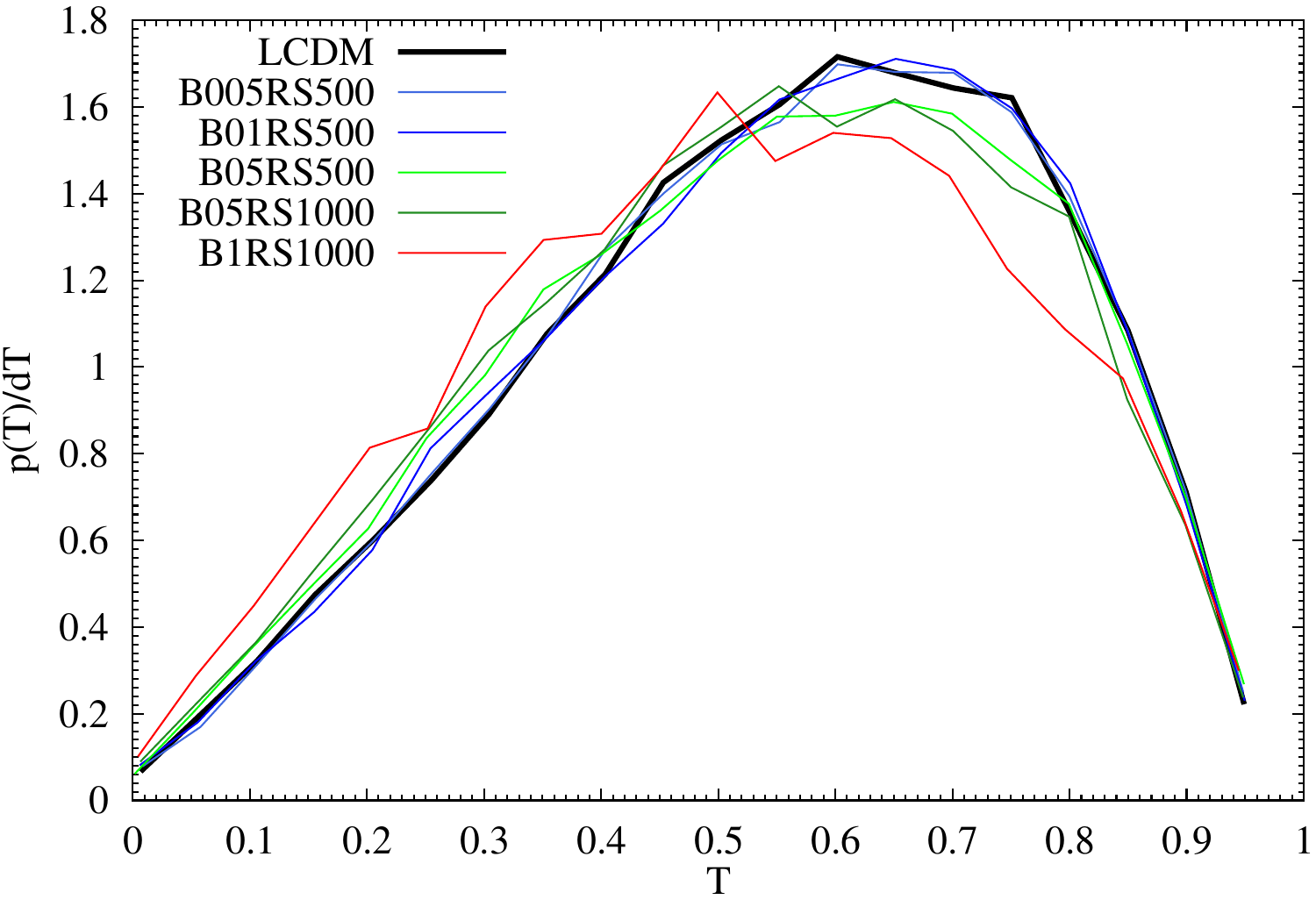}
    \caption[]{The distribution of the halo triaxiality parameter $T$ in our six models.}
    \label{fig:T_dist}
\end{figure}
\begin{figure}
        \centering
        $\begin{array}{c}
		\includegraphics[width=\figureWidthOne,angle=-90]{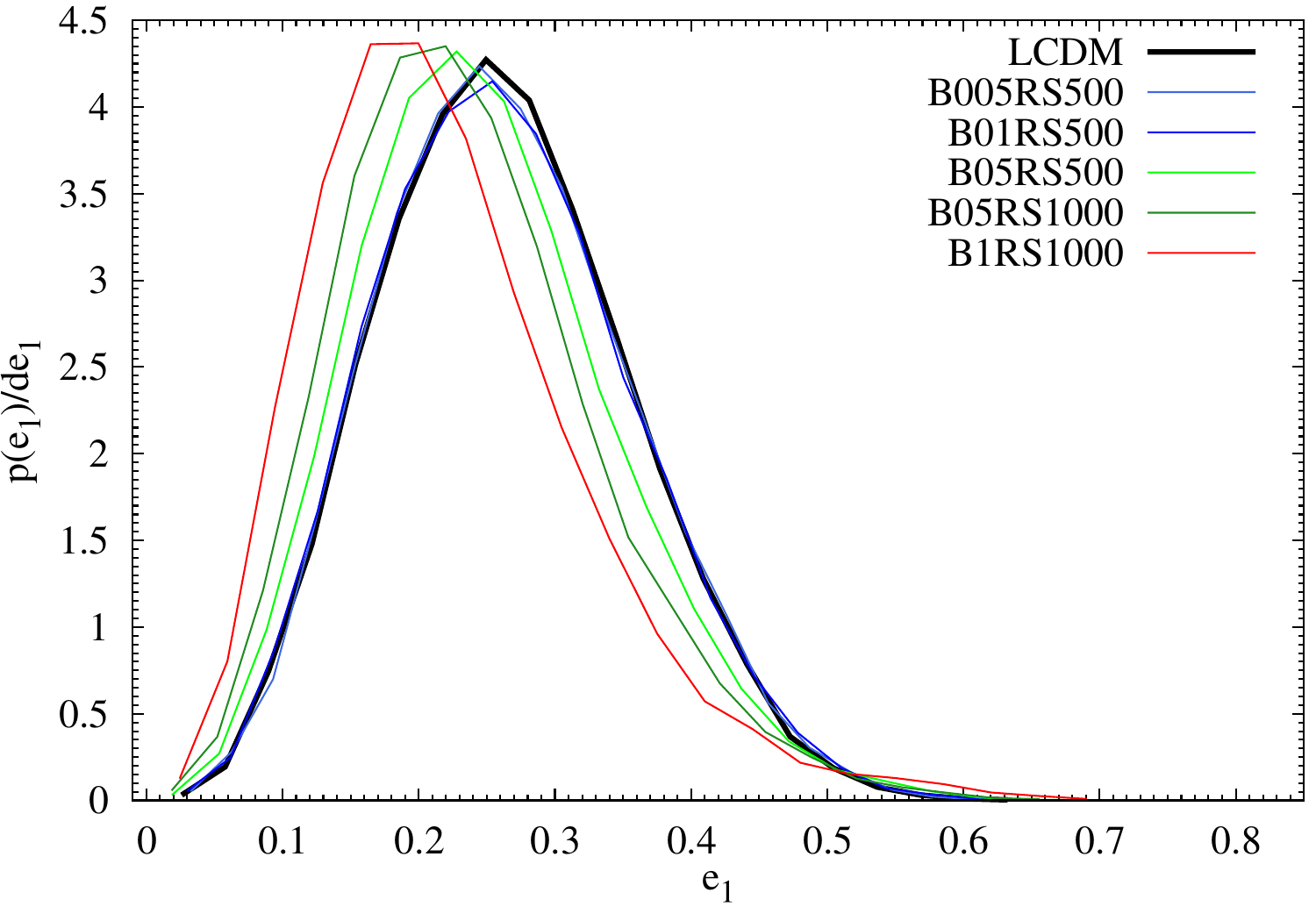} \\
		\includegraphics[width=\figureWidthOne,angle=-90]{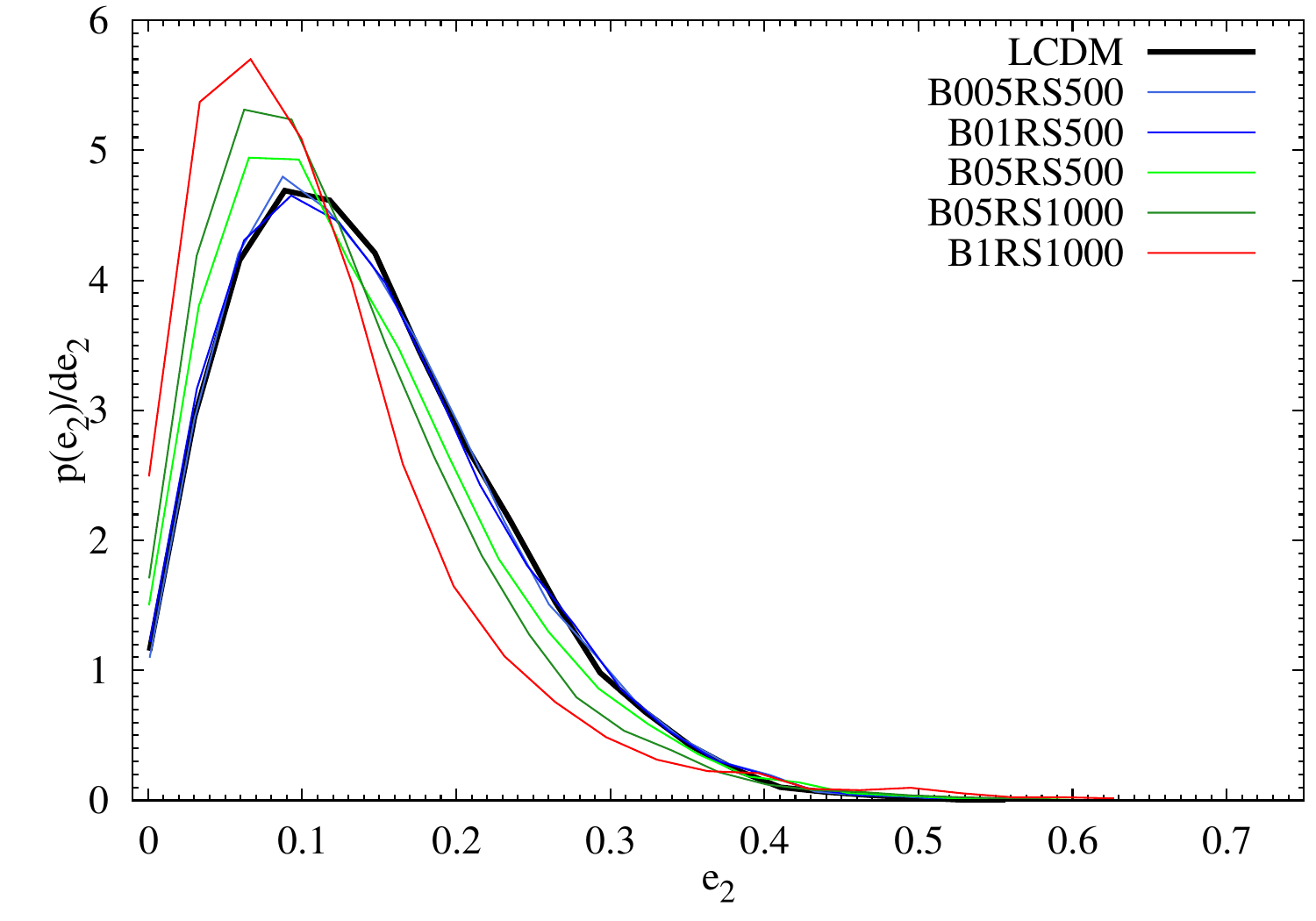}
        \end{array}$
        \caption[]{The distribution of the halo ellipticity parameter $e_1$ (top panel) and $e_2$ (bottom panel) in our six models.}
        \label{fig:e1_dist} 
\end{figure}

We determine the halo shape approximating its mass distribution to a triaxial ellipsoid. The halo's axes of inertia are calculated from the moment of inertia tensor,
which we define as:
\beq
	\label{eqn:iner_tensor}
	I_{ij} = \sum_{n}^{N_H}x_ix_j\,,
\eeq
where the particle positions $x_i$ and $x_j$ are with respect to the centre of mass of the halo and the sum covers all particles that belong to that given halo
identified by the \verb$AHF$.
The axes of the ellipsoid are found using the eigenvalues $\lambda_i$ of the inertia tensor via:
\beqa
	\label{eqn:ellpi_axies}
	a=\sqrt{\lambda_1}\,,\nonumber\\
	b=\sqrt{\lambda_2}\,,\\
	c=\sqrt{\lambda_3}\,,\nonumber
\eeqa
with $a>b>c$. As our main measure of the halo shape we use the triaxiality parameter \cite{Triaxial}:
\beq
	\label{eqn:triaxiality}
	T={a^2-b^2\over a^2-c^2}\,.
\eeq
High values of $T$ mark a prolate ellipsoid, while low values correspond to an oblate halo. There are two additional parameters related to the triaxiality,
namely the ellipticity parameters $e_1$ and $e_2$, which are defined as:
\beq
	\label{eqn:ellip}
	e_1=1-{c\over a}\quad\textrm{and}\quad e_2=1-{b\over a}\,.
\eeq
The higher the values of these parameters the less spherical is the ellipsoid's projection in the planes of the related semi-axis.

\begin{table*}[ht]
    \centering
    \caption{The mean and the standard deviations of the shape parameter distributions for all our models. }
    \label{tab:geo}
    \begin{center}
    \begin{tabular}{lllllll}
        \hline
        Model & $\langle T\rangle$ & $\sigma_T$ & $\langle e_1\rangle$ & $\sigma_{e_1}$ &  $\langle e_2\rangle$ & $\sigma_{e_2}$\\
        \hline
        LCDM      & 0.581 & 0.211 & 0.279 & 0.091 & 0.156 & 0.086\\
        B005RS500 & 0.582 & 0.211 & 0.279 & 0.092 & 0.156 & 0.087\\
        B01RS500  & 0.582 & 0.212 & 0.279 & 0.093 & 0.157 & 0.088\\
        B05RS500  & 0.572 & 0.216 & 0.266 & 0.093 & 0.148 & 0.089\\
        B05RS1000 & 0.564 & 0.215 & 0.253 & 0.094 & 0.139 & 0.086\\
        B1RS1000  & 0.547 & 0.221 & 0.239 & 0.100 & 0.129 & 0.091\\
        \hline
    \end{tabular}
    \end{center}
\end{table*}

The distributions of the shape parameters $T$, $e_1$ and $e_2$ for our whole halo population is shown in Figures~\ref{fig:T_dist} and \ref{fig:e1_dist}. These
figures illustrate that: (i) haloes for models B005RS500 and B01RS500 have distributions of the shape parameters in very good agreement with the \lcdm case;
(ii) on the other hand remaining \rebel models with a much higher $\beta$ parameter show significant departures from the control sample; (iii) measured differences
are stronger for the ellipticity parameters rather than for the triaxiality. The distributions of the \rebel shape parameters are shifted towards values indicating more spherical shapes.

To outline this claim, we collected in \reftab{tab:geo} the mean values of the shape parameters alongside
the distribution scatter $\sigma$ for all six runs.
The values in the table highlight that indeed the distributions of the shape parameters for models \lcdm, B005RS500 and B01RS500 are nearly indistinguishable.
The haloes in the models with stronger scalar forces have shapes noticeably different. For the strongest model - B1RS1000 - we find mean $T$ lower by $6\%$, $e_1$ lower
by $15\%$ and $e_2$ by $18\%$ compared to the \lcdm case. This effect can have pronounced consequence for clusters and the conundrum of the degree of anisotropy of their
velocity dispersions. Due to the limited volume of the simulation we cannot quantify properly the shape changes for clusters. Nonetheless the more spherical shapes of
\rebel haloes could indicate that on averaged they are better virialised. This can be due to the gravity enhancing effect of the scalar interactions that, in turn,
promote earlier structure formation and faster dynamical relaxation.

\subsection{Virialisation}

To investigate the virialisation state of our haloes we use the the Virial Theorem. A halo is virialised when:
\beq
	\label{eqn:vir_cond1}
	2K+U=0\;.
\eeq

However the above assumes that a halo is in complete isolation and that all the mass connected to the halo has been taken into account.
The {\tt AHF} code we use to identify haloes makes a cut at the outer boundary which corresponds to the radius at which the spherically averaged density is $\Delta_{200}= 200\times\rho_{crit}$. Particles bound to the halo, but outside this boundary, are not used to compute the overall kinetic and potential energies.
However these particles do affect the virial state of the halo and can be accounted for in the form of the pressure term $E_s$.
We compute this term for each halo individually using the method described in\cite{PressureTerm2}.

Using \eq{eqn:vir_cond1} we define the virial ratio parameter (now including the pressure term)
$\mathcal{V}$, which is a measure of the virial state of an object (for a more detailed and elaborate discussion about a halo's virial state see e.g. \cite{Power2011}):
\beq
	\label{eqn:vir_fac}
	\mathcal{V}={2K - E_s\over |U|}\,.
\eeq
The virial ratio approaches unity for a fully virialised halo, it is in range $1<\mathcal{V}\leq2$ for gravitationally bound system while $\mathcal{V}>2$ depicts an unbound object.
\begin{figure}
        \centering
		\includegraphics[angle=-90,width=\figureWidthTwo]{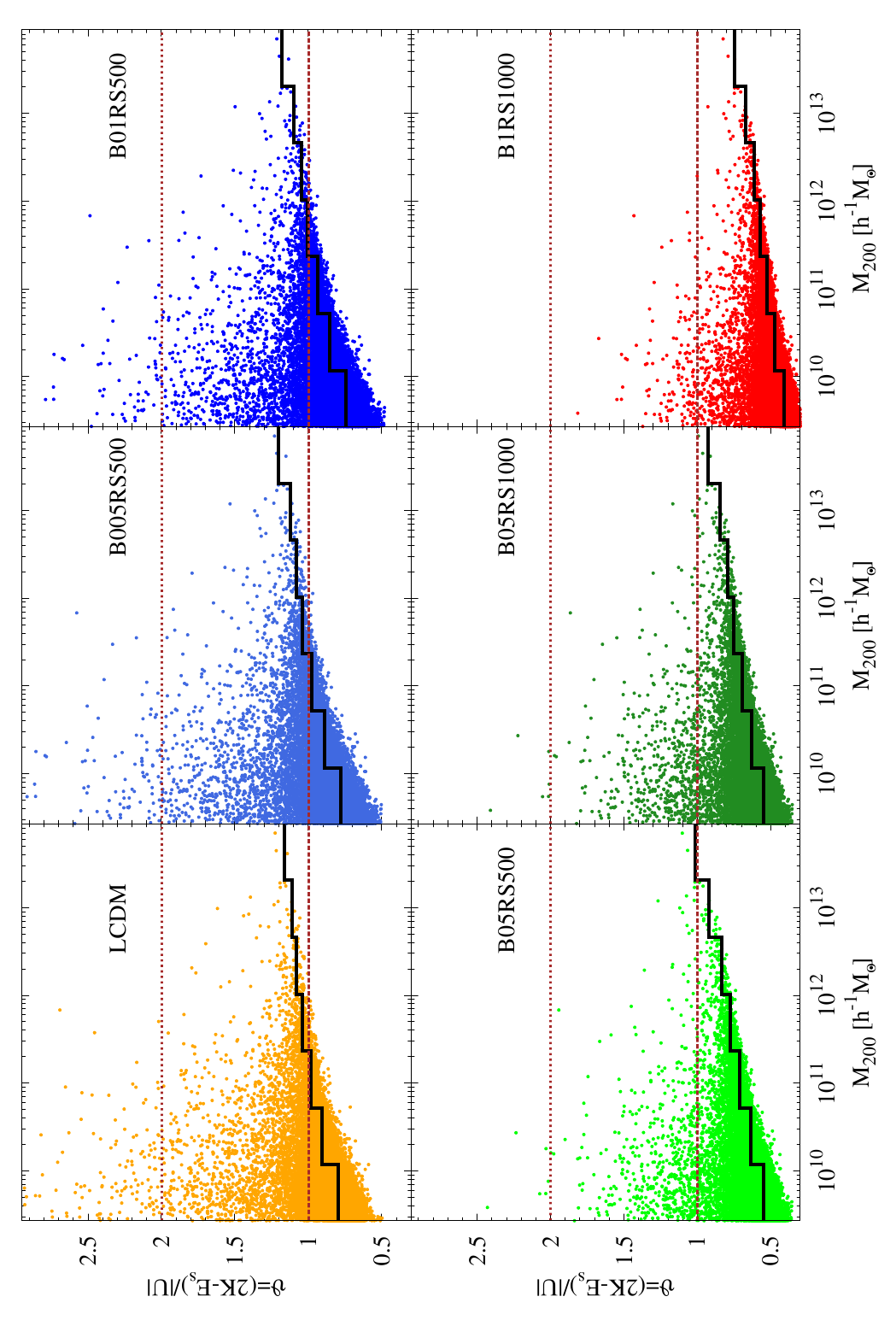}
        \caption[]{The virial ratios for haloes in all models. Panels show results from the \lcdm (top-left) to B1RS1000 (bottom-right). In each panel the points
        show results for individual haloes while the solid black lines depict the median values computed in each mass bin. Horizontal lines mark the virialisation
        threshold $\mathcal{V}=1$ (the dashed line) and the gravitational bound threshold $\mathcal{V}=2$ (the dotted line).}
        \label{fig:vir_fac}
\end{figure}

The \rebel gravitational potential (see \eq{eqn:rebel-pot}) suggest that halos with $R_{200}\ll r_s$ have a gravitational bounding energy larger by factor of $1+\beta$,
effectively lowering the virialisation factor. However the forces that particles exert on each other are also increased providing to some extent higher accelerations,
thus also higher velocity dispersions. These two effects have opposite contributions to the virialisation parameter $\mathcal{V}$. Nevertheless we can expect that on average
haloes are more virialised in the \rebel case.

\begin{figure}
        \centering
		\includegraphics[width=\figureWidthOne,angle=-90]{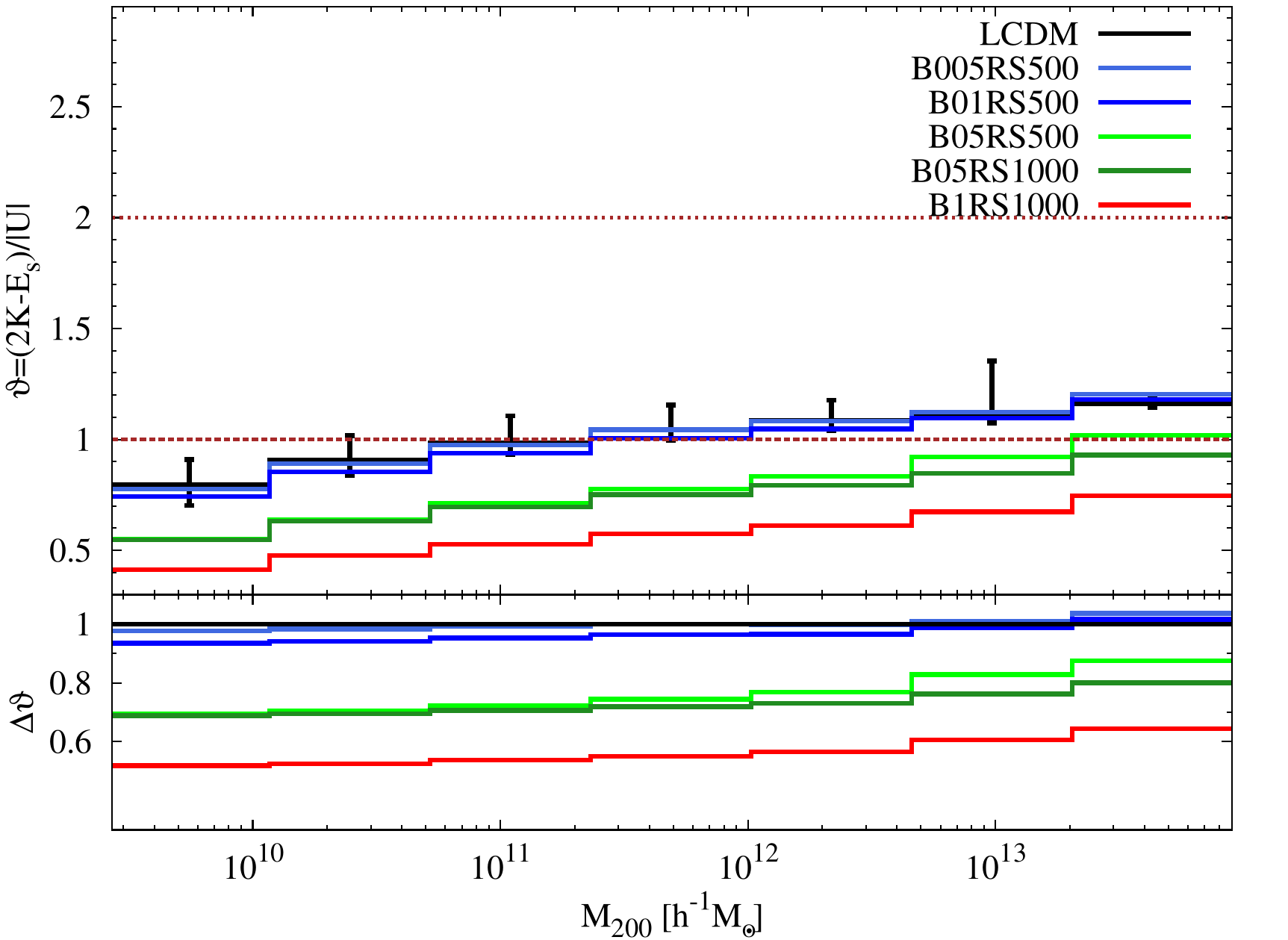}
        \caption[]{Top panel: the median virial ratios for all models binned in halo mass. The error bars represent the upper and lower quartiles of the sample and are
        plotted only for the \lcdm case. Bottom panel: The deviation $\Delta\mathcal{V}={\mathcal{V}_{ReBEL}/\mathcal{V}_{LCDM}}$ of \rebel virial ratios $\mathcal{V}_{ReBEL}$
        from the \lcdm case $\mathcal{V}_{LCDM}$.}
        \label{fig:vir_fac_comp}
\end{figure}

The virial ratios for both \lcdm and \rebel models are presented in Figures~\ref{fig:vir_fac} and \ref{fig:vir_fac_comp}. The former gives in the six panels the virial ratio
as defined by \eq{eqn:vir_fac} as a function of halo mass for each of our simulation runs. Dots represent the $\mathcal{V}$ values for individual haloes while the solid
black lines depict median virial ratio binned in halo mass. Each panel shows two brown horizontal lines marking the two virial ratio thresholds: a virialised state
(the dashed line) and a gravitationally bounded state (the dotted line). The results presented in the figure clearly underline known effects, that on average smaller
mass haloes tend to be closer to relaxation while massive ones ($M_H\simgt 10^{13}-10^{14}h^{-1}M_{\odot}$) are less virialised systems. We immediately notice that in all
our \rebel runs the majority of DM haloes have lower values of the virial state $\mathcal{V}$ than haloes in the fiducial \lcdm run. This is clearly visible when looking
at the lines depicting the averaged virial ratios binned in halo mass.

To study this effect in more detail we show in the top panel of \reffig{fig:vir_fac_comp} the median values for the virial ratios. In the bottom panel of the same figure
we plot the relative deviations of this quantity compared to the \lcdm case $\Delta\mathcal{V}=\mathcal{V}_{ReBEL}/\mathcal{V}_{LCDM}$. Undoubtedly the \rebel haloes,
on average, are more virialised than their \lcdm cousins.
For models with $\beta\geq 0.5$ the effect is very prominent and amounts to $30\%-50\%$ lower virial ratios. For the other two weaker fifth force models we find that
the overall virial ratio is very close to the \lcdm case, although systematically shifted towards lower values. \reffig{fig:vir_fac_comp} also illustrates how the
screening length affects the virial state of a halo. This can be seen when comparing the B05RS500 and B05RS1000 runs which have the same strength of the scalar forces $\beta=0.5$,
but differ in the screening length with $r_s=500\hkpc$ and respectively $r_s=1000\hkpc$. While these two models have the same median virial ratio for objects with $M\simlt 2\times 10^{11}\Msol$, the run with a larger $r_s$ has the massive haloes in a more virialised state.

\section{Conclusions}
\label{sec:conclusions}

This work studied the differences between $\lcdm$ and a specific implementation of exotic physics, the \rebel model. The new physics considered here involves long-range
fifth-forces between DM particles that phenomenologically act as modified gravity. These forces are restricted to scales below $1\hmpc$ due to the screening of
the scalar field mediating this interaction. The \rebel model is characterized by two free parameters: the ratio of the scalar to the gravitational force $\beta$
and the screening length $r_s$. In this work we analysed five such models with different values of these two parameters to explore a range of allowed \rebel cosmologies.

To asses the effects of the fifth force we conducted a series of high resolution N-body simulations that followed the formation and evolution of structure in both
the \lcdm and \rebel models. Using the results of these simulations, we have focused our attention on understanding and quantifying how halo internal properties
change as a function of the strength of the scalar interaction. We summarise our findings as follows:
\begin{itemize}
    \item The density profiles of \rebel haloes are well described by the NFW profile \cite{NFW}, but with higher concentrations. The increase in concentration
    ranges from a few percent for the weakest of our fifth-force model up to $300\%$ ($5\sigma$ away from the \lcdm mean)
    for the case with the strongest scalar interaction. This result puts the high
    $\beta$ \rebel models at odds with current astronomical observations which favour smaller halo concentrations (e.g. see\cite{dm_con1,dm_con2,dm_con3,dm_con4,dm_con5,dm_con6,
    dm_con7,dm_con8,dm_con9}). However, we must stress out that on this scales there are additional baryonic physics effects, which we did not include, that can change the picture significantly. Therefore, we need additional simulations with baryons in order to make a more realistic comparison between \rebel predictions for the halo concentration and current observational data.

    \item The analysis of the halo spins distributions revealed that \rebel haloes with $M_{200}\geq 10^{11}\Msol$ are characterised by a higher mean spin
    $\lambda_0$ and standard deviations values. This indicates that DM haloes in fifth force models have higher rotational support and thus they spin faster.
    On the other hand, we did not find any significant boost to the spin of the low-mass haloes. The spin-acquiring mechanism is enhanced by \rebel forces only
    in the regime of galaxy and cluster like haloes. This leads to the breaking of the \lcdm weak spin-mass dependence, as the high mass haloes spin faster in ReBEL.
    The variation of the mean spin with mass is especially significant for the strongest scalar interaction models that we tested - B05RS100 and B1RS1000.

    \item Our studies showed that halo shapes are sensitive to the presence of a fifth force. For strong \rebel models we found strong deviations of the triaxiality
    and especially ellipticity parameters from \lcdm. Haloes in these models tend to be more spherical, which is a direct outcome of the fact that smaller mass haloes
    ($M_{200}\leq 5\times 10^{11}\Msol$) are more virialised in \rebel than in \lcdm. For weak scalar interacting models, $\beta\leq 0.1$, we find that the halo shape
    parameters are insensitive to the presence of this additional force.

    \item The comparison of the halo virialisation state in \lcdm versus \rebel showed that the scalar forces, in average, help haloes attain dynamical
    relaxation. This effect is only prominent for strong fifth forces. The models with $\beta\leq0.1$ showed only a minor virialisation increase and only
    for objects considerably smaller than the screening length $r_s=500\hkpc$.

%
\end{itemize}

Our findings give an interesting picture for \rebel and similar fifth-force cosmologies. For the models in which the fifth force is small compared to gravity, we
found that the differences in the measured halo properties are small. As we increase the strength of the fifth force, most of the halo properties, both as distributions and as mean values, start to deviate significantly
from the \lcdm case. But the clear and prominent signature of the scalar force in the DM haloes properties is obtained only for unrealistic values of the $\beta$ parameter.
Such strong scalar forces are incompatible with the observed universe due to noticeably higher power spectrum and two-point correlation function. Moreover, we underline
that the enhanced structure formation produces haloes with higher concentration parameters. This affects especially the dwarf halo regime ($M_{200}<10^{10}\Msol$) which
puts \rebel in conflict with the observations of the Local Universe, which are already a problem for \lcdm \cite{lcdm_mw1,lcdm_mw2,lcdm_mw3}. However, we
must stress that the severe effects seen in the profiles of small-mass \rebel haloes do not necessarily trouble other modified gravity models (e.g. \cite{scalars5,Coupled_DE2,Coupled_DE5,Coupled_DE3,fofr_3,symm1}). Therefore, these effects may be a specific
feature of the \rebel cosmology where the fifth force is allowed to act starting with very early times of cosmic evolution.

Our results suggest that \rebel cosmologies with large values of the $\beta$ parameter make predictions that are at odds with current observational data, more so than the standard
\lcdm picture. The models with weak scalar interactions make similar prediction to the standard scenario and therefore it may be difficult to distinguish between the two.
The general trend of changes induced by the \rebel interactions shows that this class of modified gravity models provides at most a poor fix to the galaxy-scale challenges of
the \lcdm universe. Thus, we conclude that, in the light of this research, the \rebel cosmology is no longer an interesting and appealing modification of the \lcdm paradigm as it used to be.

\acknowledgments
Shortly after completing the first version of this paper our co-author and friend - Roman Juszkiewicz - had passed away.
Filled with grief and muse we dedicate this work to his living memory. As we are grateful for Roman's many ideas
and encouragements. We would also like to thank the anonymous referee, whose comments and suggestions helped us
significantly improve the scientific quality of this paper. The simulations used in this research were performed
on the \verb#Boreasz# supercomputer at the Interdisciplinary Center for Mathematical and computational Modelling
of the University of Warsaw. This research was carried out with the support of the "HPC Infrastructure for Grand
Challenges of Science and Engineering" Project, co-financed by the European Regional Development Fund under
the Innovative Economy Operational Programme. WAH would like to acknowledge the hospitality of the Institute of Astronomy at
 the University of Zielona G\'ora that he has received during his stay there.  WAH also acknowledge the support
of this resreach received from Polish National Science Center in grant no. DEC-2011/01/D/ST9/01960.
AK acknowledges support by the Spanish Ministerio de Ciencia e Innovacion (MICINN) in Spain through the Ramon y Cajal programme as well as the grants
AYA 2009-13875-C03-02, AYA2009-12792-C03- 03, CSD2009-00064, CAM S2009/ESP-1496 and partial support from the European Union FP7 ITN INVISIBLES
(Marie Curie Actions, PITN-GA-2011-289442). He also thanks Burt Bacharach foralfie
SRK acknowledges support by the Ministerio de Ciencia e Innovacion (MICINN) under the Consolider-Ingenio, SyeC project CSD-2007-00050.

\bibliographystyle{JHEP}
\bibliography{rebel_haloes}

\providecommand{\href}[2]{#2}\begingroup\raggedright\begin{thebibliography}{100}

\bibitem{WMAP07}
E.~{Komatsu} and {et al.}, {\it {Seven-Year Wilkinson Microwave Anisotropy
  Probe (WMAP) Observations: Cosmological Interpretation}},  {\em ArXiv
  e-prints} (Jan., 2010) [\href{http://xxx.lanl.gov/abs/1001.4538}{{\tt
  arXiv:1001.4538}}].

\bibitem{2004PhRvD..69j3501T}
M.~{Tegmark}, M.~A. {Strauss}, M.~R. {Blanton}, K.~{Abazajian}, S.~{Dodelson},
  H.~{Sandvik}, X.~{Wang}, D.~H. {Weinberg}, I.~{Zehavi}, N.~A. {Bahcall},
  F.~{Hoyle}, D.~{Schlegel}, R.~{Scoccimarro}, M.~S. {Vogeley}, A.~{Berlind},
  T.~{Budavari}, A.~{Connolly}, D.~J. {Eisenstein}, D.~{Finkbeiner}, J.~A.
  {Frieman}, J.~E. {Gunn}, L.~{Hui}, B.~{Jain}, D.~{Johnston}, S.~{Kent},
  H.~{Lin}, R.~{Nakajima}, R.~C. {Nichol}, J.~P. {Ostriker}, A.~{Pope},
  R.~{Scranton}, U.~{Seljak}, R.~K. {Sheth}, A.~{Stebbins}, A.~S. {Szalay},
  I.~{Szapudi}, Y.~{Xu}, J.~{Annis}, J.~{Brinkmann}, S.~{Burles}, F.~J.
  {Castander}, I.~{Csabai}, J.~{Loveday}, M.~{Doi}, M.~{Fukugita},
  B.~{Gillespie}, G.~{Hennessy}, D.~W. {Hogg}, {\v Z}.~{Ivezi{\'c}}, G.~R.
  {Knapp}, D.~Q. {Lamb}, B.~C. {Lee}, R.~H. {Lupton}, T.~A. {McKay},
  P.~{Kunszt}, J.~A. {Munn}, L.~{O'Connell}, J.~{Peoples}, J.~R. {Pier},
  M.~{Richmond}, C.~{Rockosi}, D.~P. {Schneider}, C.~{Stoughton}, D.~L.
  {Tucker}, D.~E. {vanden Berk}, B.~{Yanny}, and D.~G. {York}, {\it
  {Cosmological parameters from SDSS and WMAP}},  {\em \prd} {\bf 69} (May,
  2004) 103501, [\href{http://xxx.lanl.gov/abs/astro-ph/}{{\tt astro-ph/}}].

\bibitem{2006PhRvD..74l3507T}
M.~{Tegmark}, D.~J. {Eisenstein}, M.~A. {Strauss}, D.~H. {Weinberg}, M.~R.
  {Blanton}, J.~A. {Frieman}, M.~{Fukugita}, J.~E. {Gunn}, A.~J.~S. {Hamilton},
  G.~R. {Knapp}, R.~C. {Nichol}, J.~P. {Ostriker}, N.~{Padmanabhan}, W.~J.
  {Percival}, D.~J. {Schlegel}, D.~P. {Schneider}, R.~{Scoccimarro},
  U.~{Seljak}, H.-J. {Seo}, M.~{Swanson}, A.~S. {Szalay}, M.~S. {Vogeley},
  J.~{Yoo}, I.~{Zehavi}, K.~{Abazajian}, S.~F. {Anderson}, J.~{Annis}, N.~A.
  {Bahcall}, B.~{Bassett}, A.~{Berlind}, J.~{Brinkmann}, T.~{Budavari},
  F.~{Castander}, A.~{Connolly}, I.~{Csabai}, M.~{Doi}, D.~P. {Finkbeiner},
  B.~{Gillespie}, K.~{Glazebrook}, G.~S. {Hennessy}, D.~W. {Hogg}, {\v
  Z}.~{Ivezi{\'c}}, B.~{Jain}, D.~{Johnston}, S.~{Kent}, D.~Q. {Lamb}, B.~C.
  {Lee}, H.~{Lin}, J.~{Loveday}, R.~H. {Lupton}, J.~A. {Munn}, K.~{Pan},
  C.~{Park}, J.~{Peoples}, J.~R. {Pier}, A.~{Pope}, M.~{Richmond},
  C.~{Rockosi}, R.~{Scranton}, R.~K. {Sheth}, A.~{Stebbins}, C.~{Stoughton},
  I.~{Szapudi}, D.~L. {Tucker}, D.~E. {vanden Berk}, B.~{Yanny}, and D.~G.
  {York}, {\it {Cosmological constraints from the SDSS luminous red galaxies}},
   {\em \prd} {\bf 74} (Dec., 2006) 123507,
  [\href{http://xxx.lanl.gov/abs/astro-ph/}{{\tt astro-ph/}}].

\bibitem{2005MNRAS.362..505C}
S.~{Cole}, W.~J. {Percival}, J.~A. {Peacock}, P.~{Norberg}, C.~M. {Baugh},
  C.~S. {Frenk}, I.~{Baldry}, J.~{Bland-Hawthorn}, T.~{Bridges}, R.~{Cannon},
  M.~{Colless}, C.~{Collins}, W.~{Couch}, N.~J.~G. {Cross}, G.~{Dalton}, V.~R.
  {Eke}, R.~{De Propris}, S.~P. {Driver}, G.~{Efstathiou}, R.~S. {Ellis},
  K.~{Glazebrook}, C.~{Jackson}, A.~{Jenkins}, O.~{Lahav}, I.~{Lewis},
  S.~{Lumsden}, S.~{Maddox}, D.~{Madgwick}, B.~A. {Peterson}, W.~{Sutherland},
  and K.~{Taylor}, {\it {The 2dF Galaxy Redshift Survey: power-spectrum
  analysis of the final data set and cosmological implications}},  {\em \mnras}
  {\bf 362} (Sept., 2005) 505--534,
  [\href{http://xxx.lanl.gov/abs/astro-ph/}{{\tt astro-ph/}}].

\bibitem{CDM1}
M.~{Davis}, G.~{Efstathiou}, C.~S. {Frenk}, and S.~D.~M. {White}, {\it {The
  evolution of large-scale structure in a universe dominated by cold dark
  matter}},  {\em \apj} {\bf 292} (May, 1985) 371--394.

\bibitem{CDM2}
A.~{Dekel} and J.~{Silk}, {\it {The origin of dwarf galaxies, cold dark matter,
  and biased galaxy formation}},  {\em \apj} {\bf 303} (Apr., 1986) 39--55.

\bibitem{CDM3}
V.~{Trimble}, {\it {Existence and nature of dark matter in the universe}},
  {\em \araa} {\bf 25} (1987) 425--472.

\bibitem{CDM4}
C.~S. {Frenk}, S.~D.~M. {White}, M.~{Davis}, and G.~{Efstathiou}, {\it {The
  formation of dark halos in a universe dominated by cold dark matter}},  {\em
  \apj} {\bf 327} (Apr., 1988) 507--525.

\bibitem{CDM5}
A.~{Jenkins}, C.~S. {Frenk}, F.~R. {Pearce}, P.~A. {Thomas}, J.~M. {Colberg},
  S.~D.~M. {White}, H.~M.~P. {Couchman}, J.~A. {Peacock}, G.~{Efstathiou}, and
  A.~H. {Nelson}, {\it {Evolution of Structure in Cold Dark Matter Universes}},
   {\em \apj} {\bf 499} (May, 1998) 20,
  [\href{http://xxx.lanl.gov/abs/astro-ph/}{{\tt astro-ph/}}].

\bibitem{CDM6}
N.~{Arkani-Hamed}, D.~P. {Finkbeiner}, T.~R. {Slatyer}, and N.~{Weiner}, {\it
  {A theory of dark matter}},  {\em \prd} {\bf 79} (Jan., 2009) 015014,
  [\href{http://xxx.lanl.gov/abs/0810.0713}{{\tt arXiv:0810.0713}}].

\bibitem{LSCDM1}
H.~J. {Mo} and S.~D.~M. {White}, {\it {An analytic model for the spatial
  clustering of dark matter haloes}},  {\em \mnras} {\bf 282} (Sept., 1996)
  347--361, [\href{http://xxx.lanl.gov/abs/astro-ph/}{{\tt astro-ph/}}].

\bibitem{LSCDM2}
N.~A. {Bahcall}, {\it {Large-scale structure in the universe indicated by
  galaxy clusters}},  {\em \araa} {\bf 26} (1988) 631--686.

\bibitem{LSCDM3}
R.~E. {Smith}, J.~A. {Peacock}, A.~{Jenkins}, S.~D.~M. {White}, C.~S. {Frenk},
  F.~R. {Pearce}, P.~A. {Thomas}, G.~{Efstathiou}, and H.~M.~P. {Couchman},
  {\it {Stable clustering, the halo model and non-linear cosmological power
  spectra}},  {\em \mnras} {\bf 341} (June, 2003) 1311--1332,
  [\href{http://xxx.lanl.gov/abs/astro-ph/}{{\tt astro-ph/}}].

\bibitem{LSCDM4}
D.~J. {Eisenstein} and W.~{Hu}, {\it {Power Spectra for Cold Dark Matter and
  Its Variants}},  {\em \apj} {\bf 511} (Jan., 1999) 5--15,
  [\href{http://xxx.lanl.gov/abs/astro-ph/}{{\tt astro-ph/}}].

\bibitem{LSCDM5}
C.~M. {Baugh}, S.~{Cole}, C.~S. {Frenk}, and C.~G. {Lacey}, {\it {The Epoch of
  Galaxy Formation}},  {\em \apj} {\bf 498} (May, 1998) 504,
  [\href{http://xxx.lanl.gov/abs/astro-ph/}{{\tt astro-ph/}}].

\bibitem{LSCDM6}
N.~{Kaiser} and G.~{Squires}, {\it {Mapping the dark matter with weak
  gravitational lensing}},  {\em \apj} {\bf 404} (Feb., 1993) 441--450.

\bibitem{LSCDM7}
R.~K. {Sheth} and R.~{van de Weygaert}, {\it {A hierarchy of voids: much ado
  about nothing}},  {\em \mnras} {\bf 350} (May, 2004) 517--538,
  [\href{http://xxx.lanl.gov/abs/astro-ph/}{{\tt astro-ph/}}].

\bibitem{LSCDM8}
J.~R. {Bond}, L.~{Kofman}, and D.~{Pogosyan}, {\it {How filaments of galaxies
  are woven into the cosmic web}},  {\em \nat} {\bf 380} (Apr., 1996) 603--606,
  [\href{http://xxx.lanl.gov/abs/astro-ph/}{{\tt astro-ph/}}].

\bibitem{LSCDM9}
R.~{van de Weygaert} and W.~{Schaap}, {\it {The Cosmic Web: Geometric
  Analysis}},  in {\em Data Analysis in Cosmology} (V.~J. {Mart{\'{\i}}nez},
  E.~{Saar}, E.~{Mart{\'{\i}}nez-Gonz{\'a}lez}, and M.-J.
  {Pons-Border{\'{\i}}a}, eds.), vol.~665 of {\em Lecture Notes in Physics,
  Berlin Springer Verlag}, pp.~291--413, 2009.

\bibitem{Rebel}
J.~A. {Keselman}, A.~{Nusser}, and P.~J.~E. {Peebles}, {\it {Cosmology with
  equivalence principle breaking in the dark sector}},  {\em \prd} {\bf 81}
  (Mar., 2010) 063521, [\href{http://xxx.lanl.gov/abs/0912.4177}{{\tt
  arXiv:0912.4177}}].

\bibitem{rebel_nature}
P.~J.~E. {Peebles} and A.~{Nusser}, {\it {Nearby galaxies as pointers to a
  better theory of cosmic evolution}},  {\em \nat} {\bf 465} (June, 2010)
  565--569, [\href{http://xxx.lanl.gov/abs/1001.1484}{{\tt arXiv:1001.1484}}].

\bibitem{PeeblesVoid}
P.~J.~E. {Peebles}, {\it {The Void Phenomenon}},  {\em Astrophys.~J.} {\bf 557}
  (Aug., 2001) 495--504, [\href{http://xxx.lanl.gov/abs/astro-ph/}{{\tt
  astro-ph/}}].

\bibitem{GP1}
S.~S. {Gubser} and P.~J.~E. {Peebles}, {\it {Structure formation in a
  string-inspired modification of the cold dark matter model}},  {\em \prd}
  {\bf 70} (Dec., 2004) 123510, [\href{http://xxx.lanl.gov/abs/hep-th/04}{{\tt
  hep-th/04}}].

\bibitem{GP2}
S.~S. {Gubser} and P.~J.~E. {Peebles}, {\it {Cosmology with a dynamically
  screened scalar interaction in the dark sector}},  {\em \prd} {\bf 70} (Dec.,
  2004) 123511, [\href{http://xxx.lanl.gov/abs/hep-th/04}{{\tt hep-th/04}}].

\bibitem{PeeblesIDMDE}
G.~R. {Farrar} and P.~J.~E. {Peebles}, {\it {Interacting Dark Matter and Dark
  Energy}},  {\em Astrophys.~J.} {\bf 604} (Mar., 2004) 1--11,
  [\href{http://xxx.lanl.gov/abs/astro-ph/0307316}{{\tt astro-ph/0307316}}].

\bibitem{Farrar2007}
G.~R. {Farrar} and R.~A. {Rosen}, {\it {A New Force in the Dark Sector?}},
  {\em Phys.~Rev.~Lett.} {\bf 98} (Apr., 2007) 171302,
  [\href{http://xxx.lanl.gov/abs/astro-ph/0610298}{{\tt astro-ph/0610298}}].

\bibitem{NGP}
A.~{Nusser}, S.~S. {Gubser}, and P.~J. {Peebles}, {\it {Structure formation
  with a long-range scalar dark matter interaction}},  {\em \prd} {\bf 71}
  (Apr., 2005) 083505, [\href{http://xxx.lanl.gov/abs/astro-ph/0412586}{{\tt
  astro-ph/0412586}}].

\bibitem{LRSI1}
W.~A. {Hellwing} and R.~{Juszkiewicz}, {\it {Dark matter gravitational
  clustering with a long-range scalar interaction}},  {\em \prd} {\bf 80}
  (Oct., 2009) 083522, [\href{http://xxx.lanl.gov/abs/0809.1976}{{\tt
  arXiv:0809.1976}}].

\bibitem{LRSI2}
W.~A. {Hellwing}, {\it {Galactic halos in cosmology with long-range scalar DM
  interaction}},  {\em Annalen der Physik} {\bf 19} (2010), no.~3-5 351--354,
  [\href{http://xxx.lanl.gov/abs/0911.0573}{{\tt arXiv:0911.0573}}].

\bibitem{Rebel2}
W.~A. {Hellwing}, S.~R. {Knollmann}, and A.~{Knebe}, {\it {Boosting
  hierarchical structure formation with scalar-interacting dark matter}},  {\em
  \mnras} {\bf 408} (Oct., 2010) L104--L108,
  [\href{http://xxx.lanl.gov/abs/1004.2929}{{\tt arXiv:1004.2929}}].

\bibitem{reion1}
R.~{Cen}, {\it {Cosmological reionization in LCDM models with and without a
  scalar field}},  {\em New Astronomy Review} {\bf 50} (Mar., 2006) 191--198.

\bibitem{LRSI3}
W.~A. {Hellwing}, R.~{Juszkiewicz}, and R.~{van de Weygaert}, {\it {Hierarchy
  of N-point functions in the {$\Lambda$}CDM and ReBEL cosmologies}},  {\em
  \prd} {\bf 82} (Nov., 2010) 103536,
  [\href{http://xxx.lanl.gov/abs/1008.3930}{{\tt arXiv:1008.3930}}].

\bibitem{Coupled_DE1}
M.~{Baldi}, V.~{Pettorino}, G.~{Robbers}, and V.~{Springel}, {\it
  {Hydrodynamical N-body simulations of coupled dark energy cosmologies}},
  {\em \mnras} {\bf 403} (Apr., 2010) 1684--1702,
  [\href{http://xxx.lanl.gov/abs/0812.3901}{{\tt arXiv:0812.3901}}].

\bibitem{Coupled_DE2}
M.~{Baldi}, {\it {Simulations of structure formation in interacting dark energy
  cosmologies}},  {\em Nuclear Physics B Proceedings Supplements} {\bf 194}
  (Oct., 2009) 178--184, [\href{http://xxx.lanl.gov/abs/0906.5353}{{\tt
  arXiv:0906.5353}}].

\bibitem{Coupled_DE5}
M.~{Baldi} and P.~{Salucci}, {\it {Constraints on interacting Dark Energy
  models from galaxy rotation curves}},  {\em Journal of Cosmology and
  Astro-Particle Physics} {\bf 2} (Feb., 2012) 14,
  [\href{http://xxx.lanl.gov/abs/1111.3953}{{\tt arXiv:1111.3953}}].

\bibitem{scalars1}
B.~{Li} and J.~D. {Barrow}, {\it {On the effects of coupled scalar fields on
  structure formation}},  {\em \mnras} {\bf 413} (May, 2011) 262--270,
  [\href{http://xxx.lanl.gov/abs/1010.3748}{{\tt arXiv:1010.3748}}].

\bibitem{scalars2}
B.~{Li}, {\it {Voids in coupled scalar field cosmology}},  {\em \mnras} {\bf
  411} (Mar., 2011) 2615--2627, [\href{http://xxx.lanl.gov/abs/1009.1406}{{\tt
  arXiv:1009.1406}}].

\bibitem{scalars3}
B.~{Li}, D.~F. {Mota}, and J.~D. {Barrow}, {\it {N-body Simulations for
  Extended Quintessence Models}},  {\em \apj} {\bf 728} (Feb., 2011) 109,
  [\href{http://xxx.lanl.gov/abs/1009.1400}{{\tt arXiv:1009.1400}}].

\bibitem{scalars4}
B.~{Li} and J.~D. {Barrow}, {\it {N-body simulations for coupled scalar-field
  cosmology}},  {\em \prd} {\bf 83} (Jan., 2011) 024007,
  [\href{http://xxx.lanl.gov/abs/1005.4231}{{\tt arXiv:1005.4231}}].

\bibitem{scalars5}
M.~{Baldi}, {\it {Clarifying the effects of interacting dark energy on linear
  and non-linear structure formation processes}},  {\em \mnras} {\bf 414}
  (June, 2011) 116--128, [\href{http://xxx.lanl.gov/abs/1012.0002}{{\tt
  arXiv:1012.0002}}].

\bibitem{scalars6}
M.~{Baldi}, {\it {Time-dependent couplings in the dark sector: from background
  evolution to non-linear structure formation}},  {\em \mnras} {\bf 411} (Feb.,
  2011) 1077--1103, [\href{http://xxx.lanl.gov/abs/1005.2188}{{\tt
  arXiv:1005.2188}}].

\bibitem{scalars7}
A.-C. {Davis}, B.~{Li}, D.~F. {Mota}, and H.~A. {Winther}, {\it {Structure
  Formation in the Symmetron Model}},  {\em ArXiv e-prints} (Aug., 2011)
  [\href{http://xxx.lanl.gov/abs/1108.3081}{{\tt arXiv:1108.3081}}].

\bibitem{scalars8}
M.~{Baldi}, {\it {Early massive clusters and the bouncing coupled dark
  energy}},  {\em ArXiv e-prints} (July, 2011)
  [\href{http://xxx.lanl.gov/abs/1107.5049}{{\tt arXiv:1107.5049}}].

\bibitem{scalars9}
M.~{Baldi}, J.~{Lee}, and A.~V. {Macci{\`o}}, {\it {The Effect of Coupled Dark
  Energy on the Alignment Between Dark Matter and Galaxy Distributions in
  Clusters}},  {\em \apj} {\bf 732} (May, 2011) 112,
  [\href{http://xxx.lanl.gov/abs/1101.5761}{{\tt arXiv:1101.5761}}].

\bibitem{scalars10}
B.~{Li} and H.~{Zhao}, {\it {Structure formation by the fifth force:
  Segregation of baryons and dark matter}},  {\em \prd} {\bf 81} (May, 2010)
  104047, [\href{http://xxx.lanl.gov/abs/1001.3152}{{\tt arXiv:1001.3152}}].

\bibitem{scalars11}
H.~{Zhao}, A.~V. {Macci{\`o}}, B.~{Li}, H.~{Hoekstra}, and M.~{Feix}, {\it
  {Structure Formation by Fifth Force: Power Spectrum from N-Body
  Simulations}},  {\em \apjl} {\bf 712} (Apr., 2010) L179--L183,
  [\href{http://xxx.lanl.gov/abs/0910.3207}{{\tt arXiv:0910.3207}}].

\bibitem{scalars12}
B.~{Li}, W.~A. {Hellwing}, K.~{Koyama}, G.-B. {Zhao}, E.~{Jennings}, and C.~M.
  {Baugh}, {\it {The non-linear matter and velocity power spectra in f(R)
  gravity}},  {\em \mnras} {\bf 428} (Jan., 2013) 743--755,
  [\href{http://xxx.lanl.gov/abs/1206.4317}{{\tt arXiv:1206.4317}}].

\bibitem{offset}
H.~{Shan}, B.~{Qin}, B.~{Fort}, C.~{Tao}, X.-P. {Wu}, and H.~{Zhao}, {\it
  {Offset between dark matter and ordinary matter: evidence from a sample of 38
  lensing clusters of galaxies}},  {\em \mnras} {\bf 406} (Aug., 2010)
  1134--1139, [\href{http://xxx.lanl.gov/abs/1004.1475}{{\tt
  arXiv:1004.1475}}].

\bibitem{UKIDSS_z09_supercluster}
A.~M. {Swinbank} and {et al.}, {\it {The discovery of a massive supercluster at
  z = 0.9 in the UKIDSS Deep eXtragalactic Survey}},  {\em \mnras} {\bf 379}
  (Aug., 2007) 1343--1351, [\href{http://xxx.lanl.gov/abs/0706.0090}{{\tt
  arXiv:0706.0090}}].

\bibitem{higz_cluster_obs1}
R.~J. {Foley}, K.~{Andersson}, G.~{Bazin}, T.~{de Haan}, J.~{Ruel}, P.~A.~R.
  {Ade}, K.~A. {Aird}, R.~{Armstrong}, M.~L.~N. {Ashby}, M.~{Bautz}, B.~A.
  {Benson}, L.~E. {Bleem}, M.~{Bonamente}, M.~{Brodwin}, J.~E. {Carlstrom},
  C.~L. {Chang}, A.~{Clocchiatti}, T.~M. {Crawford}, A.~T. {Crites},
  S.~{Desai}, M.~A. {Dobbs}, J.~P. {Dudley}, G.~G. {Fazio}, W.~R. {Forman},
  G.~{Garmire}, E.~M. {George}, M.~D. {Gladders}, A.~H. {Gonzalez}, N.~W.
  {Halverson}, F.~W. {High}, G.~P. {Holder}, W.~L. {Holzapfel}, S.~{Hoover},
  J.~D. {Hrubes}, C.~{Jones}, M.~{Joy}, R.~{Keisler}, L.~{Knox}, A.~T. {Lee},
  E.~M. {Leitch}, M.~{Lueker}, D.~{Luong-Van}, D.~P. {Marrone}, J.~J.
  {McMahon}, J.~{Mehl}, S.~S. {Meyer}, J.~J. {Mohr}, T.~E. {Montroy}, S.~S.
  {Murray}, S.~{Padin}, T.~{Plagge}, C.~{Pryke}, C.~L. {Reichardt}, A.~{Rest},
  J.~E. {Ruhl}, B.~R. {Saliwanchik}, A.~{Saro}, K.~K. {Schaffer}, L.~{Shaw},
  E.~{Shirokoff}, J.~{Song}, H.~G. {Spieler}, B.~{Stalder}, S.~A. {Stanford},
  Z.~{Staniszewski}, A.~A. {Stark}, K.~{Story}, C.~W. {Stubbs},
  K.~{Vanderlinde}, J.~D. {Vieira}, A.~{Vikhlinin}, R.~{Williamson}, and
  A.~{Zenteno}, {\it {Discovery and Cosmological Implications of SPT-CL
  J2106-5844, the Most Massive Known Cluster at z>1}},  {\em \apj} {\bf 731}
  (Apr., 2011) 86, [\href{http://xxx.lanl.gov/abs/1101.1286}{{\tt
  arXiv:1101.1286}}].

\bibitem{higz_cluster_obs2}
M.~{Brodwin}, J.~{Ruel}, P.~A.~R. {Ade}, K.~A. {Aird}, K.~{Andersson}, M.~L.~N.
  {Ashby}, M.~{Bautz}, G.~{Bazin}, B.~A. {Benson}, L.~E. {Bleem}, J.~E.
  {Carlstrom}, C.~L. {Chang}, T.~M. {Crawford}, A.~T. {Crites}, T.~{de Haan},
  S.~{Desai}, M.~A. {Dobbs}, J.~P. {Dudley}, G.~G. {Fazio}, R.~J. {Foley},
  W.~R. {Forman}, G.~{Garmire}, E.~M. {George}, M.~D. {Gladders}, A.~H.
  {Gonzalez}, N.~W. {Halverson}, F.~W. {High}, G.~P. {Holder}, W.~L.
  {Holzapfel}, J.~D. {Hrubes}, C.~{Jones}, M.~{Joy}, R.~{Keisler}, L.~{Knox},
  A.~T. {Lee}, E.~M. {Leitch}, M.~{Lueker}, D.~P. {Marrone}, J.~J. {McMahon},
  J.~{Mehl}, S.~S. {Meyer}, J.~J. {Mohr}, T.~E. {Montroy}, S.~S. {Murray},
  S.~{Padin}, T.~{Plagge}, C.~{Pryke}, C.~L. {Reichardt}, A.~{Rest}, J.~E.
  {Ruhl}, K.~K. {Schaffer}, L.~{Shaw}, E.~{Shirokoff}, J.~{Song}, H.~G.
  {Spieler}, B.~{Stalder}, S.~A. {Stanford}, Z.~{Staniszewski}, A.~A. {Stark},
  C.~W. {Stubbs}, K.~{Vanderlinde}, J.~D. {Vieira}, A.~{Vikhlinin},
  R.~{Williamson}, Y.~{Yang}, O.~{Zahn}, and A.~{Zenteno}, {\it {SPT-CL
  J0546-5345: A Massive z>1 Galaxy Cluster Selected Via the Sunyaev-Zel'dovich
  Effect with the South Pole Telescope}},  {\em \apj} {\bf 721} (Sept., 2010)
  90--97, [\href{http://xxx.lanl.gov/abs/1006.5639}{{\tt arXiv:1006.5639}}].

\bibitem{vectorDE}
E.~{Carlesi}, A.~{Knebe}, G.~{Yepes}, S.~{Gottloeber}, J.~{Beltran Jimenez},
  and A.~L. {Maroto}, {\it {Vector dark energy and high-z massive clusters}},
  {\em ArXiv e-prints} (Aug., 2011)
  [\href{http://xxx.lanl.gov/abs/1108.4173}{{\tt arXiv:1108.4173}}].

\bibitem{highz_clustersDE}
M.~{Baldi} and V.~{Pettorino}, {\it {High-z massive clusters as a test for
  dynamical coupled dark energy}},  {\em \mnras} {\bf 412} (Mar., 2011) L1--L5,
  [\href{http://xxx.lanl.gov/abs/1006.3761}{{\tt arXiv:1006.3761}}].

\bibitem{highz_clustersLCDM}
M.~J. {Mortonson}, W.~{Hu}, and D.~{Huterer}, {\it {Simultaneous falsification
  of {$\Lambda$}CDM and quintessence with massive, distant clusters}},  {\em
  \prd} {\bf 83} (Jan., 2011) 023015,
  [\href{http://xxx.lanl.gov/abs/1011.0004}{{\tt arXiv:1011.0004}}].

\bibitem{Nordstrom1912}
G.~{Nordstr{\"o}m}, {\it {Zur Theorie der Gravitation vom Standpunkt des
  Relativit{\"a}tsprinzips}},  {\em Annalen der Physik} {\bf 347} (1913)
  533--554.

\bibitem{Yukawa1935}
H.~{Yukawa}, {\it {On the Interaction of Elementary Particles}},  {\em
  Proceedings of the Physico-Mathematical Society of Japan} {\bf 17} (1935)
  48--57.

\bibitem{Dicke1}
R.~H. {Dicke}, {\it {Scalar-Tensor Gravitation and the Cosmic Fireball}},  {\em
  \apj} {\bf 152} (Apr., 1968) 1.

\bibitem{Dicke2}
R.~H. {Dicke}, {\it {Long-Range Scalar Interaction}},  {\em Physical Review}
  {\bf 126} (June, 1962) 1875--1877.

\bibitem{Jordan}
P.~{Jordan}, {\it {Zum gegenw{\"a}rtigen Stand der Diracschen kosmologischen
  Hypothesen}},  {\em Zeitschrift fur Physik} {\bf 157} (Feb., 1959) 112--121.

\bibitem{DGB}
T.~{Damour}, G.~W. {Gibbons}, and C.~{Gundlach}, {\it {Dark matter,
  time-varying G, and a dilaton field}},  {\em Physical Review Letters} {\bf
  64} (Jan., 1990) 123--126.

\bibitem{GradwohlFrieman1}
B.-A. {Gradwohl} and J.~A. {Frieman}, {\it {Dark matter, long-range forces, and
  large-scale structure}},  {\em Astrophys.~J.} {\bf 398} (Oct., 1992)
  407--424.

\bibitem{GradwohlFrieman2}
J.~A. {Frieman} and B.-A. {Gradwohl}, {\it {Dark matter and the equivalence
  principle}},  {\em Physical Review Letters} {\bf 67} (Nov., 1991) 2926--2929.

\bibitem{skalar_teo1}
J.~A. {Casas}, J.~{Garcia-Bellido}, and M.~{Quiros}, {\it {Scalar-tensor
  theories of gravity with Phi -dependent masses}},  {\em Classical and Quantum
  Gravity} {\bf 9} (May, 1992) 1371--1384,
  [\href{http://xxx.lanl.gov/abs/hep-ph/92}{{\tt hep-ph/92}}].

\bibitem{skalar_teo2}
T.~{Damour} and A.~M. {Polyakov}, {\it {String theory and gravity}},  {\em
  General Relativity and Gravitation} {\bf 26} (Dec., 1994) 1171--1176,
  [\href{http://xxx.lanl.gov/abs/gr-qc/941}{{\tt gr-qc/941}}].

\bibitem{skalar_teo3}
C.~{Wetterich}, {\it {An asymptotically vanishing time-dependent cosmological
  ''constant''.}},  {\em \aap} {\bf 301} (Sept., 1995) 321,
  [\href{http://xxx.lanl.gov/abs/hep-th/94}{{\tt hep-th/94}}].

\bibitem{skalar_teo4}
G.~W. {Anderson} and S.~M. {Carroll}, {\it {Dark Matter with Time-Dependent
  Mass}},  in {\em COSMO-97, First International Workshop on Particle Physics
  and the Early Universe} ({L.~Roszkowski}, ed.), p.~227, 1998.

\bibitem{skalar_teo5}
R.~{Bean}, {\it {Perturbation evolution with a nonminimally coupled scalar
  field}},  {\em \prd} {\bf 64} (Dec., 2001) 123516,
  [\href{http://xxx.lanl.gov/abs/astro-ph/0104464}{{\tt astro-ph/0104464}}].

\bibitem{skalar_teo6}
L.~{Amendola}, {\it {Perturbations in a coupled scalar field cosmology}},  {\em
  \mnras} {\bf 312} (Mar., 2000) 521--530,
  [\href{http://xxx.lanl.gov/abs/astro-ph/9906073}{{\tt astro-ph/9906073}}].

\bibitem{skalar_teo7}
L.~{Amendola} and D.~{Tocchini-Valentini}, {\it {Baryon bias and structure
  formation in an accelerating universe}},  {\em \prd} {\bf 66} (Aug., 2002)
  043528, [\href{http://xxx.lanl.gov/abs/astro-ph/0111535}{{\tt
  astro-ph/0111535}}].

\bibitem{skalar_teo8}
U.~{Fran{\c c}a} and R.~{Rosenfeld}, {\it {Fine Tuning in Quintessence Models
  with Exponential Potentials}},  {\em Journal of High Energy Physics} {\bf 10}
  (Oct., 2002) 15, [\href{http://xxx.lanl.gov/abs/astro-ph/0206194}{{\tt
  astro-ph/0206194}}].

\bibitem{skalar_teo9}
T.~{Damour}, F.~{Piazza}, and G.~{Veneziano}, {\it {Violations of the
  equivalence principle in a dilaton-runaway scenario}},  {\em \prd} {\bf 66}
  (Aug., 2002) 046007, [\href{http://xxx.lanl.gov/abs/hep-th/02}{{\tt
  hep-th/02}}].

\bibitem{skalar_teo10}
D.~{Comelli}, M.~{Pietroni}, and A.~{Riotto}, {\it {Dark energy and dark
  matter}},  {\em Physics Letters B} {\bf 571} (Oct., 2003) 115--120,
  [\href{http://xxx.lanl.gov/abs/hep-ph/03}{{\tt hep-ph/03}}].

\bibitem{skalar_teo11}
L.~{Amendola}, M.~{Gasperini}, and F.~{Piazza}, {\it {Fitting type Ia
  supernovae with coupled dark energy}},  {\em Journal of Cosmology and
  Astro-Particle Physics} {\bf 9} (Sept., 2004) 14,
  [\href{http://xxx.lanl.gov/abs/astro-ph/0407573}{{\tt astro-ph/0407573}}].

\bibitem{multic-dm1}
A.~W. {Brookfield}, C.~{van de Bruck}, and L.~M.~H. {Hall}, {\it {New
  interactions in the dark sector mediated by dark energy}},  {\em \prd} {\bf
  77} (Feb., 2008) 043006, [\href{http://xxx.lanl.gov/abs/0709.2297}{{\tt
  arXiv:0709.2297}}].

\bibitem{multic-dm2}
M.~{Baldi}, {\it {Structure formation in multiple dark matter cosmologies with
  long-range scalar interactions}},  {\em \mnras} {\bf 428} (Jan., 2013)
  2074--2084, [\href{http://xxx.lanl.gov/abs/1206.2348}{{\tt
  arXiv:1206.2348}}].

\bibitem{multic-dm3}
M.~{Baldi}, {\it {Cosmological models with multiple dark matter species and
  long-range scalar interactions}},  {\em ArXiv e-prints} (Apr., 2013)
  [\href{http://xxx.lanl.gov/abs/1304.5178}{{\tt arXiv:1304.5178}}].

\bibitem{multic-dm4}
M.~{Baldi}, {\it {Multiple dark matter as a self-regulating mechanism for dark
  sector interactions}},  {\em Annalen der Physik} {\bf 524} (Oct., 2012)
  602--617, [\href{http://xxx.lanl.gov/abs/1204.0514}{{\tt arXiv:1204.0514}}].

\bibitem{1980Peebles}
P.~J.~E. {Peebles}, {\em {The large-scale structure of the universe}}.
\newblock Research supported by the National Science Foundation.~Princeton,
  N.J., Princeton University Press, 1980.~435 p., 1980.

\bibitem{fofr_2}
W.~A. {Hellwing}, B.~{Li}, C.~S. {Frenk}, and S.~{Cole}, {\it {Hierarchical
  clustering in chameleon \$f(R)\$ gravity}},  {\em ArXiv e-prints} (May, 2013)
  [\href{http://xxx.lanl.gov/abs/1305.7486}{{\tt arXiv:1305.7486}}].

\bibitem{Gadget2}
V.~{Springel}, {\it {The cosmological simulation code GADGET-2}},  {\em \mnras}
  {\bf 364} (Dec., 2005) 1105--1134,
  [\href{http://xxx.lanl.gov/abs/astro-ph/0505010}{{\tt astro-ph/0505010}}].

\bibitem{MHF}
S.~P.~D. {Gill}, A.~{Knebe}, and B.~K. {Gibson}, {\it {The evolution of
  substructure - I. A new identification method}},  {\em \mnras} {\bf 351}
  (June, 2004) 399--409, [\href{http://xxx.lanl.gov/abs/astro-ph/0404258}{{\tt
  astro-ph/0404258}}].

\bibitem{AHF}
S.~R. {Knollmann} and A.~{Knebe}, {\it {AHF: Amiga's Halo Finder}},  {\em
  \apjs} {\bf 182} (June, 2009) 608--624,
  [\href{http://xxx.lanl.gov/abs/0904.3662}{{\tt arXiv:0904.3662}}].

\bibitem{bar1}
S.~D.~M. {White} and M.~J. {Rees}, {\it {Core condensation in heavy halos - A
  two-stage theory for galaxy formation and clustering}},  {\em \mnras} {\bf
  183} (May, 1978) 341--358.

\bibitem{bar2}
D.~J. {Croton}, V.~{Springel}, S.~D.~M. {White}, G.~{De Lucia}, C.~S. {Frenk},
  L.~{Gao}, A.~{Jenkins}, G.~{Kauffmann}, J.~F. {Navarro}, and N.~{Yoshida},
  {\it {The many lives of active galactic nuclei: cooling flows, black holes
  and the luminosities and colours of galaxies}},  {\em \mnras} {\bf 365}
  (Jan., 2006) 11--28, [\href{http://xxx.lanl.gov/abs/astro-ph/}{{\tt
  astro-ph/}}].

\bibitem{bar3}
S.~D.~M. {White} and C.~S. {Frenk}, {\it {Galaxy formation through hierarchical
  clustering}},  {\em \apj} {\bf 379} (Sept., 1991) 52--79.

\bibitem{bar4}
G.~R. {Blumenthal}, S.~M. {Faber}, R.~{Flores}, and J.~R. {Primack}, {\it
  {Contraction of dark matter galactic halos due to baryonic infall}},  {\em
  \apj} {\bf 301} (Feb., 1986) 27--34.

\bibitem{bar5}
D.~H. {Rudd}, A.~R. {Zentner}, and A.~V. {Kravtsov}, {\it {Effects of Baryons
  and Dissipation on the Matter Power Spectrum}},  {\em \apj} {\bf 672} (Jan.,
  2008) 19--32, [\href{http://xxx.lanl.gov/abs/astro-ph/}{{\tt astro-ph/}}].

\bibitem{bar6}
T.~{Sawala}, C.~S. {Frenk}, R.~A. {Crain}, A.~{Jenkins}, J.~{Schaye},
  T.~{Theuns}, and J.~{Zavala}, {\it {The abundance of (not just) dark matter
  haloes}},  {\em \mnras} {\bf 431} (May, 2013) 1366--1382,
  [\href{http://xxx.lanl.gov/abs/1206.6495}{{\tt arXiv:1206.6495}}].

\bibitem{equiv1}
M.~{Kesden} and M.~{Kamionkowski}, {\it {Tidal tails test the equivalence
  principle in the dark-matter sector}},  {\em \prd} {\bf 74} (Oct., 2006)
  083007, [\href{http://xxx.lanl.gov/abs/astro-ph/}{{\tt astro-ph/}}].

\bibitem{equiv2}
J.~A. {Keselman}, A.~{Nusser}, and P.~J.~E. {Peebles}, {\it {Galaxy satellites
  and the weak equivalence principle}},  {\em \prd} {\bf 80} (Sept., 2009)
  063517, [\href{http://xxx.lanl.gov/abs/0902.3452}{{\tt arXiv:0902.3452}}].

\bibitem{mog_gadget}
E.~{Puchwein}, M.~{Baldi}, and V.~{Springel}, {\it {Modified Gravity-GADGET: A
  new code for cosmological hydrodynamical simulations of modified gravity
  models}},  {\em ArXiv e-prints} (May, 2013)
  [\href{http://xxx.lanl.gov/abs/1305.2418}{{\tt arXiv:1305.2418}}].

\bibitem{NFW}
J.~F. {Navarro}, C.~S. {Frenk}, and S.~D.~M. {White}, {\it {A Universal Density
  Profile from Hierarchical Clustering}},  {\em \apj} {\bf 490} (Dec., 1997)
  493, [\href{http://xxx.lanl.gov/abs/astro-ph/9611107}{{\tt
  astro-ph/9611107}}].

\bibitem{Wechsler2002}
R.~H. {Wechsler}, J.~S. {Bullock}, J.~R. {Primack}, A.~V. {Kravtsov}, and
  A.~{Dekel}, {\it {Concentrations of Dark Halos from Their Assembly
  Histories}},  {\em Astrophys.~J.} {\bf 568} (Mar., 2002) 52--70,
  [\href{http://xxx.lanl.gov/abs/astro-ph/0108151}{{\tt astro-ph/0108151}}].

\bibitem{lcdm_mw1}
M.~{Boylan-Kolchin}, J.~S. {Bullock}, and M.~{Kaplinghat}, {\it {The Milky
  Way's bright satellites as an apparent failure of {$\Lambda$}CDM}},  {\em
  \mnras} {\bf 422} (May, 2012) 1203--1218,
  [\href{http://xxx.lanl.gov/abs/1111.2048}{{\tt arXiv:1111.2048}}].

\bibitem{lcdm_mw2}
M.~{Boylan-Kolchin}, J.~S. {Bullock}, and M.~{Kaplinghat}, {\it {Too big to
  fail? The puzzling darkness of massive Milky Way subhaloes}},  {\em \mnras}
  {\bf 415} (July, 2011) L40--L44,
  [\href{http://xxx.lanl.gov/abs/1103.0007}{{\tt arXiv:1103.0007}}].

\bibitem{lcdm_mw3}
J.~{Wang}, C.~S. {Frenk}, J.~F. {Navarro}, L.~{Gao}, and T.~{Sawala}, {\it {The
  missing massive satellites of the Milky Way}},  {\em \mnras} {\bf 424} (Aug.,
  2012) 2715--2721, [\href{http://xxx.lanl.gov/abs/1203.4097}{{\tt
  arXiv:1203.4097}}].

\bibitem{amom1}
F.~{Hoyle}, {\it {The Origin of the Rotations of the Galaxies}},  in {\em
  Problems of Cosmical Aerodynamics}, p.~195, 1951.

\bibitem{amom2}
P.~J.~E. {Peebles}, {\it {Origin of the Angular Momentum of Galaxies}},  {\em
  \apj} {\bf 155} (Feb., 1969) 393.

\bibitem{amom3}
A.~G. {Doroshkevich}, {\it {The space structure of perturbations and the origin
  of rotation of galaxies in the theory of fluctuation.}},  {\em Astrofizika}
  {\bf 6} (1970) 581--600.

\bibitem{amom4}
G.~{Efstathiou} and B.~J.~T. {Jones}, {\it {The rotation of galaxies -
  Numerical investigations of the tidal torque theory}},  {\em \mnras} {\bf
  186} (Jan., 1979) 133--144.

\bibitem{amom5}
S.~M. {Fall} and G.~{Efstathiou}, {\it {Formation and rotation of disc galaxies
  with haloes}},  {\em \mnras} {\bf 193} (Oct., 1980) 189--206.

\bibitem{amom6}
S.~D.~M. {White}, {\it {Angular momentum growth in protogalaxies}},  {\em \apj}
  {\bf 286} (Nov., 1984) 38--41.

\bibitem{Angular1}
P.~J.~E. {Peebles}, {\it {Origin of the Angular Momentum of Galaxies}},  {\em
  \apj} {\bf 155} (Feb., 1969) 393.

\bibitem{Angular2}
M.~{Vitvitska}, A.~A. {Klypin}, A.~V. {Kravtsov}, R.~H. {Wechsler}, J.~R.
  {Primack}, and J.~S. {Bullock}, {\it {The Origin of Angular Momentum in Dark
  Matter Halos}},  {\em \apj} {\bf 581} (Dec., 2002) 799--809,
  [\href{http://xxx.lanl.gov/abs/astro-ph/}{{\tt astro-ph/}}].

\bibitem{Angular3}
J.~{Barnes} and G.~{Efstathiou}, {\it {Angular momentum from tidal torques}},
  {\em \apj} {\bf 319} (Aug., 1987) 575--600.

\bibitem{Angular4}
M.~S. {Warren}, P.~J. {Quinn}, J.~K. {Salmon}, and W.~H. {Zurek}, {\it {Dark
  halos formed via dissipationless collapse. I - Shapes and alignment of
  angular momentum}},  {\em \apj} {\bf 399} (Nov., 1992) 405--425.

\bibitem{Angular5}
B.~{Robertson}, J.~S. {Bullock}, T.~J. {Cox}, T.~{Di Matteo}, L.~{Hernquist},
  V.~{Springel}, and N.~{Yoshida}, {\it {A Merger-driven Scenario for
  Cosmological Disk Galaxy Formation}},  {\em \apj} {\bf 645} (July, 2006)
  986--1000, [\href{http://xxx.lanl.gov/abs/astro-ph/}{{\tt astro-ph/}}].

\bibitem{Bullock2001}
J.~S. {Bullock}, A.~{Dekel}, T.~S. {Kolatt}, A.~V. {Kravtsov}, A.~A. {Klypin},
  C.~{Porciani}, and J.~R. {Primack}, {\it {A Universal Angular Momentum
  Profile for Galactic Halos}},  {\em Astrophys.~J.} {\bf 555} (July, 2001)
  240--257, [\href{http://xxx.lanl.gov/abs/astro-ph/0011001}{{\tt
  astro-ph/0011001}}].

\bibitem{spin1}
M.~S. {Warren}, P.~J. {Quinn}, J.~K. {Salmon}, and W.~H. {Zurek}, {\it {Dark
  halos formed via dissipationless collapse. I - Shapes and alignment of
  angular momentum}},  {\em \apj} {\bf 399} (Nov., 1992) 405--425.

\bibitem{spin2}
S.~{Cole} and C.~{Lacey}, {\it {The structure of dark matter haloes in
  hierarchical clustering models}},  {\em \mnras} {\bf 281} (July, 1996) 716,
  [\href{http://xxx.lanl.gov/abs/astro-ph/9510147}{{\tt astro-ph/9510147}}].

\bibitem{spin3}
H.~J. {Mo}, S.~{Mao}, and S.~D.~M. {White}, {\it {The formation of galactic
  discs}},  {\em \mnras} {\bf 295} (Apr., 1998) 319--336,
  [\href{http://xxx.lanl.gov/abs/astro-ph/9707093}{{\tt astro-ph/9707093}}].

\bibitem{spin4}
M.~{Steinmetz} and M.~{Bartelmann}, {\it {On the spin parameter of dark-matter
  haloes}},  {\em \mnras} {\bf 272} (Feb., 1995) 570--578,
  [\href{http://xxx.lanl.gov/abs/astro-ph/9403017}{{\tt astro-ph/9403017}}].

\bibitem{spin5}
P.~{Catelan} and T.~{Theuns}, {\it {Evolution of the angular momentum of
  protogalaxies from tidal torques: Zel'dovich approximation}},  {\em \mnras}
  {\bf 282} (Sept., 1996) 436--454,
  [\href{http://xxx.lanl.gov/abs/astro-ph/9604077}{{\tt astro-ph/9604077}}].

\bibitem{disk_halo1}
F.~{Governato}, B.~{Willman}, L.~{Mayer}, A.~{Brooks}, G.~{Stinson},
  O.~{Valenzuela}, J.~{Wadsley}, and T.~{Quinn}, {\it {Forming disc galaxies in
  {$\Lambda$}CDM simulations}},  {\em \mnras} {\bf 374} (Feb., 2007)
  1479--1494, [\href{http://xxx.lanl.gov/abs/astro-ph/0602351}{{\tt
  astro-ph/0602351}}].

\bibitem{disk_halo2}
B.~{Robertson}, J.~S. {Bullock}, T.~J. {Cox}, T.~{Di Matteo}, L.~{Hernquist},
  V.~{Springel}, and N.~{Yoshida}, {\it {A Merger-driven Scenario for
  Cosmological Disk Galaxy Formation}},  {\em \apj} {\bf 645} (July, 2006)
  986--1000, [\href{http://xxx.lanl.gov/abs/astro-ph/0503369}{{\tt
  astro-ph/0503369}}].

\bibitem{spin_mass1}
A.~V. {Macci{\`o}}, A.~A. {Dutton}, F.~C. {van den Bosch}, B.~{Moore},
  D.~{Potter}, and J.~{Stadel}, {\it {Concentration, spin and shape of dark
  matter haloes: scatter and the dependence on mass and environment}},  {\em
  \mnras} {\bf 378} (June, 2007) 55--71,
  [\href{http://xxx.lanl.gov/abs/astro-ph/0608157}{{\tt astro-ph/0608157}}].

\bibitem{spin_mass2}
P.~{Bett}, V.~{Eke}, C.~S. {Frenk}, A.~{Jenkins}, J.~{Helly}, and J.~{Navarro},
  {\it {The spin and shape of dark matter haloes in the Millennium simulation
  of a {$\Lambda$} cold dark matter universe}},  {\em \mnras} {\bf 376} (Mar.,
  2007) 215--232, [\href{http://xxx.lanl.gov/abs/astro-ph/0608607}{{\tt
  astro-ph/0608607}}].

\bibitem{spin_mass3}
A.~{Knebe} and C.~{Power}, {\it {On the Correlation between Spin Parameter and
  Halo Mass}},  {\em \apj} {\bf 678} (May, 2008) 621--626,
  [\href{http://xxx.lanl.gov/abs/0801.4453}{{\tt arXiv:0801.4453}}].

\bibitem{geo1}
J.~M. {Bardeen}, J.~R. {Bond}, N.~{Kaiser}, and A.~S. {Szalay}, {\it {The
  statistics of peaks of Gaussian random fields}},  {\em \apj} {\bf 304} (May,
  1986) 15--61.

\bibitem{geo2}
J.~R. {Bond} and S.~T. {Myers}, {\it {The Peak-Patch Picture of Cosmic
  Catalogs. I. Algorithms}},  {\em \apjs} {\bf 103} (Mar., 1996) 1.

\bibitem{geo3}
R.~{van de Weygaert} and E.~{Bertschinger}, {\it {Peak and gravity constraints
  in Gaussian primordial density fields: An application of the Hoffman-Ribak
  method}},  {\em \mnras} {\bf 281} (July, 1996) 84,
  [\href{http://xxx.lanl.gov/abs/astro-ph/9507024}{{\tt astro-ph/9507024}}].

\bibitem{geo4}
R.~K. {Sheth}, H.~J. {Mo}, and G.~{Tormen}, {\it {Ellipsoidal collapse and an
  improved model for the number and spatial distribution of dark matter
  haloes}},  {\em \mnras} {\bf 323} (May, 2001) 1--12,
  [\href{http://xxx.lanl.gov/abs/astro-ph/9907024}{{\tt astro-ph/9907024}}].

\bibitem{geo5}
J.~R. {Bond}, L.~{Kofman}, and D.~{Pogosyan}, {\it {How filaments of galaxies
  are woven into the cosmic web}},  {\em \nat} {\bf 380} (Apr., 1996) 603--606,
  [\href{http://xxx.lanl.gov/abs/astro-ph/9512141}{{\tt astro-ph/9512141}}].

\bibitem{geo6}
R.~{van de Weygaert}, {\it {Clusters and the Cosmic Web}},  {\em ArXiv
  Astrophysics e-prints} (July, 2006)
  [\href{http://xxx.lanl.gov/abs/astro-ph/0607539}{{\tt astro-ph/0607539}}].

\bibitem{geo7}
M.~{van Haarlem} and R.~{van de Weygaert}, {\it {Velocity Fields and Alignments
  of Clusters in Gravitational Instability Scenarios}},  {\em \apj} {\bf 418}
  (Dec., 1993) 544.

\bibitem{Triaxial}
M.~{Franx}, G.~{Illingworth}, and T.~{de Zeeuw}, {\it {The ordered nature of
  elliptical galaxies - Implications for their intrinsic angular momenta and
  shapes}},  {\em \apj} {\bf 383} (Dec., 1991) 112--134.

\bibitem{PressureTerm2}
L.~D. {Shaw}, J.~{Weller}, J.~P. {Ostriker}, and P.~{Bode}, {\it {Statistics of
  Physical Properties of Dark Matter Clusters}},  {\em \apj} {\bf 646} (Aug.,
  2006) 815--833, [\href{http://xxx.lanl.gov/abs/astro-ph/}{{\tt astro-ph/}}].

\bibitem{Power2011}
C.~{Power}, A.~{Knebe}, and S.~R. {Knollmann}, {\it {The dynamical state of
  dark matter haloes in cosmological simulations - I. Correlations with mass
  assembly history}},  {\em \mnras} (Oct., 2011) 1734,
  [\href{http://xxx.lanl.gov/abs/1109.2671}{{\tt arXiv:1109.2671}}].

\bibitem{dm_con1}
G.~{Gentile}, P.~{Salucci}, U.~{Klein}, D.~{Vergani}, and P.~{Kalberla}, {\it
  {The cored distribution of dark matter in spiral galaxies}},  {\em \mnras}
  {\bf 351} (July, 2004) 903--922,
  [\href{http://xxx.lanl.gov/abs/astro-ph/}{{\tt astro-ph/}}].

\bibitem{dm_con2}
T.~{Broadhurst}, N.~{Ben{\'{\i}}tez}, D.~{Coe}, K.~{Sharon}, K.~{Zekser},
  R.~{White}, H.~{Ford}, R.~{Bouwens}, J.~{Blakeslee}, M.~{Clampin},
  N.~{Cross}, M.~{Franx}, B.~{Frye}, G.~{Hartig}, G.~{Illingworth},
  L.~{Infante}, F.~{Menanteau}, G.~{Meurer}, M.~{Postman}, D.~R. {Ardila},
  F.~{Bartko}, R.~A. {Brown}, C.~J. {Burrows}, E.~S. {Cheng}, P.~D. {Feldman},
  D.~A. {Golimowski}, T.~{Goto}, C.~{Gronwall}, D.~{Herranz}, B.~{Holden},
  N.~{Homeier}, J.~E. {Krist}, M.~P. {Lesser}, A.~R. {Martel}, G.~K. {Miley},
  P.~{Rosati}, M.~{Sirianni}, W.~B. {Sparks}, S.~{Steindling}, H.~D. {Tran},
  Z.~I. {Tsvetanov}, and W.~{Zheng}, {\it {Strong-Lensing Analysis of A1689
  from Deep Advanced Camera Images}},  {\em \apj} {\bf 621} (Mar., 2005)
  53--88, [\href{http://xxx.lanl.gov/abs/astro-ph/}{{\tt astro-ph/}}].

\bibitem{dm_con3}
J.~D. {Simon}, A.~D. {Bolatto}, A.~{Leroy}, L.~{Blitz}, and E.~L. {Gates}, {\it
  {High-Resolution Measurements of the Halos of Four Dark Matter-Dominated
  Galaxies: Deviations from a Universal Density Profile}},  {\em \apj} {\bf
  621} (Mar., 2005) 757--776, [\href{http://xxx.lanl.gov/abs/astro-ph/}{{\tt
  astro-ph/}}].

\bibitem{dm_con4}
J.~{Diemand}, M.~{Zemp}, B.~{Moore}, J.~{Stadel}, and C.~M. {Carollo}, {\it
  {Cusps in cold dark matter haloes}},  {\em \mnras} {\bf 364} (Dec., 2005)
  665--673, [\href{http://xxx.lanl.gov/abs/astro-ph/}{{\tt astro-ph/}}].

\bibitem{dm_con5}
L.~E. {Strigari}, J.~S. {Bullock}, M.~{Kaplinghat}, J.~{Diemand}, M.~{Kuhlen},
  and P.~{Madau}, {\it {Redefining the Missing Satellites Problem}},  {\em
  \apj} {\bf 669} (Nov., 2007) 676--683,
  [\href{http://xxx.lanl.gov/abs/0704.1817}{{\tt arXiv:0704.1817}}].

\bibitem{dm_con6}
G.~{Gentile}, A.~{Burkert}, P.~{Salucci}, U.~{Klein}, and F.~{Walter}, {\it
  {The Dwarf Galaxy DDO 47 as a Dark Matter Laboratory: Testing Cusps Hiding in
  Triaxial Halos}},  {\em \apjl} {\bf 634} (Dec., 2005) L145--L148,
  [\href{http://xxx.lanl.gov/abs/astro-ph/}{{\tt astro-ph/}}].

\bibitem{dm_con7}
E.~{Hayashi}, J.~F. {Navarro}, C.~{Power}, A.~{Jenkins}, C.~S. {Frenk},
  S.~D.~M. {White}, V.~{Springel}, J.~{Stadel}, and T.~R. {Quinn}, {\it {The
  inner structure of {$\Lambda$}CDM haloes - II. Halo mass profiles and low
  surface brightness galaxy rotation curves}},  {\em \mnras} {\bf 355} (Dec.,
  2004) 794--812, [\href{http://xxx.lanl.gov/abs/astro-ph/}{{\tt astro-ph/}}].

\bibitem{dm_con8}
A.~A. {El-Zant}, Y.~{Hoffman}, J.~{Primack}, F.~{Combes}, and I.~{Shlosman},
  {\it {Flat-cored Dark Matter in Cuspy Clusters of Galaxies}},  {\em \apjl}
  {\bf 607} (June, 2004) L75--L78,
  [\href{http://xxx.lanl.gov/abs/astro-ph/}{{\tt astro-ph/}}].

\bibitem{dm_con9}
J.~{Diemand}, M.~{Kuhlen}, P.~{Madau}, M.~{Zemp}, B.~{Moore}, D.~{Potter}, and
  J.~{Stadel}, {\it {Clumps and streams in the local dark matter
  distribution}},  {\em \nat} {\bf 454} (Aug., 2008) 735--738,
  [\href{http://xxx.lanl.gov/abs/0805.1244}{{\tt arXiv:0805.1244}}].

\bibitem{Coupled_DE3}
B.~{Li} and J.~D. {Barrow}, {\it {N-Body Simulations for Coupled Scalar Field
  Cosmology}},  {\em ArXiv e-prints} (May, 2010)
  [\href{http://xxx.lanl.gov/abs/1005.4231}{{\tt arXiv:1005.4231}}].

\bibitem{fofr_3}
B.~{Li}, G.-B. {Zhao}, and K.~{Koyama}, {\it {Haloes and voids in f(R)
  gravity}},  {\em \mnras} {\bf 421} (Apr., 2012) 3481--3487,
  [\href{http://xxx.lanl.gov/abs/1111.2602}{{\tt arXiv:1111.2602}}].

\bibitem{symm1}
H.~A. {Winther}, D.~F. {Mota}, and B.~{Li}, {\it {Environment Dependence of
  Dark Matter Halos in Symmetron Modified Gravity}},  {\em \apj} {\bf 756}
  (Sept., 2012) 166, [\href{http://xxx.lanl.gov/abs/1110.6438}{{\tt
  arXiv:1110.6438}}].

\end{thebibliography}\endgroup

\end{document}